\documentclass[aps,prl,superscriptaddress,twocolumn,floatfix]{revtex4-2}
\usepackage[abs]{overpic}
\usepackage{graphicx,graphics}
\usepackage{dcolumn}
\usepackage{amsmath,amssymb,amsfonts}
\usepackage{wasysym}
\usepackage{latexsym,verbatim}
\usepackage{bm}
\usepackage[dvipsnames]{xcolor}
\usepackage[framemethod=default]{mdframed}
\usepackage[breaklinks=true,colorlinks,citecolor=blue,linkcolor=blue,urlcolor=blue]{hyperref}
\usepackage[makeroom]{cancel}
\usepackage{fancybox}
\usepackage{ulem}
\usepackage{balance}

\newcommand{\be}{\begin{eqnarray}}
\newcommand{\ee}{\end{eqnarray}}
\newcommand{\nn}{\nonumber\\}

\newcommand{\sgn}{\text{sgn}}

\begin{document}

\title{No bulk thermal currents in massive Dirac fermions}

\author{Feng Liu}
\email{feng.liu-6@postgrad.manchester.ac.uk}
\author{A. Daria Dumitriu-I.}
\email{alexandra-daria.dumitriu-iovanescu@outlook.com}
\author{Alessandro Principi}%
\email{alessandro.principi@manchester.ac.uk}
\affiliation{%
Department of Physics and Astronomy, University of Manchester, M13 9PL Manchester (UK)
 }%


\begin{abstract}
We calculate the energy current flowing in the bulk of a (2+1)-dimensional system of massive Dirac fermions and along a (1+1)-dimensional domain wall generated by flipping the sign of the particle mass. We show that, at low temperatures and in the long-wavelenghth limit, the system does not support a bulk thermal Hall current proportional to the temperature gradient. The only such contribution is due to states localised at the domain wall. This puts an end to a controversy existing in the literature and amends previous results obtained via first-order perturbation calculations.

\end{abstract}

\maketitle

%
{\it Introduction}---The thermal Hall effect mirrors the Hall effect in the realm of energy transport, {\it i.e.} a transverse heat current emerges when a temperature gradient is established.
Such effect has been studied in a variety of systems, such as multiferroics~\cite{ideue2017giant}, topological insulators~\cite{shimizu2015quantum}, magnets~\cite{katsura2010theory,zhuo2021topological,kawano2019thermal,owerre2017topological}, quantum spin liquids~\cite{kasahara2018unusual,kasahara2018majorana,teng2020unquantized}, strongly-correlated systems~\cite{uehara2022phonon,auerbach2019equilibrium} to name a few, and it also constitutes an important diagnostics of neutral excitations. From a theoretical standpoint, thermal responses can be calculated by using  Luttinger's ``trick''~\cite{luttinger1964theory}, by exploiting the equivalency that exists in the linear-response regime between a non-flat metric tensor (also termed ``gravitational potential'' in what follows, in analogy with the electric potential) and temperature fluctuations in generating energy currents~\cite{luttinger1964theory,cooper1997thermoelectric,qin2011energy}. In this scheme, the Hamiltonian is perturbed by introducing a non-trivial metric tensor that couples to the system's energy density. The first derivative of the metric tensor then defines the gravitational field (analogous to the electric field and equivalent to the thermal gradient within linear response) that generates the longitudinal and/or Hall thermal currents~\cite{luttinger1964theory,cooper1997thermoelectric,qin2011energy}.

Recently, a controversy has arisen about whether thermal Hall currents proportional to the  gravitational field ({\it i.e.} to the temperature gradient) can be supported in the system's bulk~\cite{shitade2014heat,nakai2016finite,nakai2017laughlin}, or whether such currents are {\it always} proportional to higher-order gradients of the temperature fluctuations~\cite{stone2012gravitational,bradlyn2015low,gromov2015thermal,vinkler2019bulk,huang2022torsion}. Refs.~\cite{stone2012gravitational,bradlyn2015low} were the first to point out a fundamental difference between the charge and thermal Hall effects. They showed that, in contrast to transverse charge currents that arise in response to uniform electric fields, within a gravitational Chern-Simons action no thermal Hall current is generated in response to a uniform gravitational field (temperature gradient) in the bulk of a (2+1)-dimensional gapped system. This result was challenged by Refs.~\cite{shitade2014heat,nakai2016finite,nakai2017laughlin}. In particular, Ref.~\cite{nakai2016finite} found a bulk contribution proportional to the first derivative of the gravitational potential, when calculating the thermal Hall response to first order in the metric tensor and in the long-wavelength limit.

More recent analytical and numerical works~\cite{gromov2015thermal,vinkler2019bulk,huang2022torsion} are in agreement with the earlier findings of Refs.~\cite{stone2012gravitational,bradlyn2015low}. However, in deriving their results, Refs.~\cite{gromov2015thermal,vinkler2019bulk,huang2022torsion} do not follow the methods employed in~\cite{shitade2014heat,nakai2016finite,nakai2017laughlin}. It is therefore unclear whether these latter works present shortcomings that can be remedied to get results consistent with the rest of the literature~\cite{stone2012gravitational,bradlyn2015low,gromov2015thermal,vinkler2019bulk,huang2022torsion}. The scope of this work is thus to put an end to such controversy by showing, with an analytical calculation that closely follows that of Ref.~\cite{nakai2016finite}, that it is indeed possible to obtain the correct results by including all-order contributions in the metric tensor to the system's free energy. By doing so, we show that the bulk thermal Hall response proportional to a uniform gravitational field (temperature gradient) vanishes. Thus, the thermal Hall response in the long-wavelength limit features only boundary contributions.

\begin{figure}[t!]
    \centering
    \includegraphics{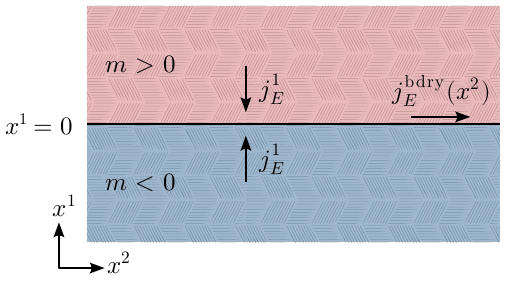}
    \caption{The boundary located at $x^1=0$ separates two (2+1)-dimensional bulk massive-Dirac-Fermion systems, whose masses are equal but of opposite sign: negative in the half-plane $x^{1}<0$ and positive in the half-plane $x^{1}>0$. A boundary current $j^{\text{bdry}}_E$ flows along the edge, {\it i.e.} in the $x^{2}$ direction, and bulk currents $j^{1}_E$ flow across the boundary. These currents satisfy the continuity equation given in Eq.~\eqref{eq:bulk_current}. The textured background represents a non-zero gravitational potential.} 
    \label{fig:one}
\end{figure}

In this paper we consider the system of Fig.~\ref{fig:one}, {\it i.e.} a (1+1)-dimensional boundary located at $x^1=0$ and oriented along the direction $x^2$ which separates two (2+1)-dimensional bulk massive-Dirac-Fermion systems. We assume their masses in the two half-spaces, $x^1>0$ and $x^1<0$, to be equal in magnitude but opposite in sign, as shown in Fig.~\ref{fig:one}. The continuity of energy currents imposes that a bulk thermal Hall current must necessarily exist to account for boundary anomalies. This is to say that, if a (1+1)-dimensional current $j^{\text{bdry}}_E$ flows along the boundary, then the bulk current $j^{1}_E$ flowing {\it across} the boundary must satisfy the continuity equation~\footnote{Note that the boundary energy current is not dependent on time, since the boundary free energy~(\ref{eq:free_energy_result}) and responses derived from it are static. Eq.~(\ref{eq:bulk_current}) is obtained by assuming that a steady state is achieved. Therefore, the divergence of the energy current must be equal to zero. At the boundary, the current has two contributions: the bulk one $j^{1}_E(x^1,x^2)$ flowing in the $x^1$ direction, and the boundary one $j^{2}_E(x^1,x^2) = j^{\text{bdry}}_E(x^2) \delta(x^1)$ flowing along the boundary in the direction $x^2$ and localized at $x^1=0$. Integrating the equation $\partial_\mu j^{\mu}_E = 0$ over $x^1$ across the boundary and using that, by symmetry, $j^{1}_E(x^1=-0)=-j^1_E(x^1=+0)$, one readily obtains Eq.~(\ref{eq:bulk_current}).}
\be \label{eq:bulk_current}
j^{1}_E(x^1=-0)=-j^1_E(x^1=+0)=\frac{1}{2}\partial_2j^{\text{bdry}}_E.
\ee
This equation is central in what follows in proving that the bulk current $j^{1}_E(x^1)$ can only be proportional to the derivative of the gravitational field ({\it i.e.} to the second derivative of the temperature fluctuations). 

On the contrary, the calculation of Ref.~\cite{nakai2016finite} suggests that the boundary current is directly proportional to the metric tensor. If this were true, according to Eq.~(\ref{eq:bulk_current}), the bulk thermal Hall current would be proportional to the first derivative of the metric, and therefore to the temperature gradient. However, these findings are derived from lowest-order approximations of the quantities involved: higher-order corrections could dramatically alter this conclusion. 

To show that this is indeed the case, we follow the method used in Ref.~\cite{nakai2016finite} but we calculate contributions to the thermal current to {\it all} orders in the gravitational potential in the long-wavelength limit. We omit terms in the boundary current that depend on the derivatives of the metric, since such terms would correspond to contributions to the bulk thermal Hall current proportional to at least the second derivative of the gravitational potential. Our all-order calculation shows that the long-wavelength energy current flowing along a boundary located at $x^1=0$ is
\be \label{eq:boundary_current_result}
j_E^{\text{bdry}}(x^2)
=\frac{\pi T^2}{12}
\ee
{\it i.e.} it depends only on the uniform equilibrium temperature $T$ and does not contain any term proportional to the gravitational potential itself ({\it i.e.} to the temperature fluctuations away from equilibrium). According to the thermal generalization of the Streda formula~\cite{nomura2012cross}, we can use the result of Eq.~(\ref{eq:boundary_current_result}) to find the thermal Hall conductivity, $\kappa_{xy} = \sgn(m)\pi T/12$.~\cite{bradlyn2015low,nakai2016finite,nomura2012cross}.

Eq.~(\ref{eq:boundary_current_result}) implies that $\partial_2j_E^{\text{bdry}} = 0$ and therefore one can immediately conclude that there is no bulk thermal Hall current which is proportional to the gravitational field (temperature gradient). Thus, by following the same method presented in Ref.~\cite{nakai2016finite}, our all-order calculation corrects their approximate result. Going beyond the long-wavelength approximation,  one would include contributions to the boundary energy current of Eq.~(\ref{eq:boundary_current_result}) that are proportional to the first derivative of the metric tensor. These in turn translate, via Eq.~(\ref{eq:bulk_current}), into leading-order contributions to the bulk thermal Hall current that are proportional to the second derivative of the temperature fluctuations away from equilibrium. This result is in agreement with the effective theories of Refs.~\cite{stone2012gravitational,gromov2015thermal} and numerical results of Ref.~\cite{vinkler2019bulk}.

{\it The model}---The action for $(2+1)$-dimensional Dirac fermions coupled to a gravitational field is (hereafter, we set $\hbar = 1$)~\footnote{In this paper, we have adopted a form of the action that differs from the one in Ref.~\cite{nakai2016finite}. Specifically, we have added total derivatives to the Lagrangian to make it Hermitian. However, this difference does not alter our final conclusions. Based on the action from Refs.~\cite{nakai2016finite}, the boundary free energy obtained is the same as the one presented in Eq.~(\ref{eq:free_energy_result}) of the main text, up to a constant that is independent of temperature.} 
\be \label{eq:action_def}
S=
\int_{x,t} \sqrt{g} \bar\psi \left[
\frac{i}{2}(e_{\phantom{.}\alpha}^{\mu}\gamma^{\alpha} 
\overrightarrow{\nabla}_{\mu} - \overleftarrow{\nabla}_{\mu}\gamma^{\alpha}e_{\phantom{.}\alpha}^{\mu})-m
\right]\psi,
\ee
where $\int_{x,t}=\int d^2x dt$ and
the covariant derivative $\overrightarrow{\nabla}_{\mu}$ ($\overleftarrow{\nabla}_{\mu}$) acts on the right (left) two-component spinor field $\psi$. Explicitly, $\overrightarrow{\nabla}_{\mu}\psi = \overrightarrow{\partial}_{\mu}\psi +[\gamma_{\alpha},\gamma_{\beta}]\omega_{\mu}^{\phantom{.}\alpha\beta} \psi/8$, where $\overrightarrow{\partial}_\mu$ is the derivative over the temporal ($\mu=0$) and spatial ($\mu = x,y$) directions, while $\omega_{\mu}^{\phantom{.}\alpha\beta} =e_{\nu}^{\phantom{.}\alpha}e^{\nu'}_{\phantom{.}\beta}\Gamma^{\nu}_{\mu\nu'}-e^{\nu}_{\phantom{.}\beta}\partial_{\mu}e_{\nu}^{\phantom{.}\alpha}$ is the 
spin connection \cite{carroll2019spacetime}. Finally, the combinations  $\gamma^0\gamma^1$, $\gamma^0\gamma^2$ and $\gamma^0$ correspond to the usual Pauli matrices $\sigma_x$, $\sigma_y$ and $\sigma_z$, respectively. In  Eq.~(\ref{eq:action_def}), we have introduced the metric $g_{\mu\nu}$,  whose determinant, in modulus, is $g$. The factor $\sqrt{g}$ ensures invariance of the action under changes of coordinates. Throughout this paper, we use the Greek indices $\mu,\nu=0,1,2$ and $\alpha,\beta,\ldots =\hat{0},\hat{1},\hat{2}$ to denote the environment and locally flat (or internal) coordinates, respectively. In what follows, when we refer to space-like directions only, we will use the Latin letters $i,j=1,2$ for the environment coordinates, and $a,b=\hat{1},\hat{2}$ for the internal coordinates. The Minkowski metric in the locally-flat space-time is taken to be $\eta_{\alpha\beta}=\text{diag}(+1,-1,-1)$. The environment and flat metrics, $g_{\mu\nu}$ and $\eta_{\alpha\beta}$, respectively, are related by a vielbein field $e^{\phantom{.}\alpha}_{\mu}$ according to the identity~\cite{carroll2019spacetime}
$g_{\mu\nu}=e^{\phantom{.}\alpha}_{\mu}e^{\phantom{.}\beta}_{\nu}\eta_{\alpha\beta}$.
From Eq.~(\ref{eq:action_def}), we define the energy-momentum tensor
\be
\tau^{\mu}_{\phantom{.}\nu}=e^{\phantom{.}\alpha}_{\nu}\tau^{\mu}_{\phantom{.}\alpha}=
-e^{\phantom{.}\alpha}_{\nu}\frac{1}{\sqrt{g}}\frac{\delta S}{\delta e_{\mu}^{\phantom{.}\alpha}},
\ee
and the Hamiltonian acting in an effectively locally-flat space-time as 
\be \label{eq:local_Hamiltonian}
H&=&\int_x \sqrt{g}\bar\psi\Big[\frac{i}{2}e^{0}_{\phantom{.}\alpha}\gamma^\alpha\omega_0+\frac{i}{2}\omega_0\gamma^\alpha e^{0}_{\phantom{.}\alpha}
\nn
&&\qquad-\frac{i}{2}e^{j}_{\phantom{.}\alpha}\gamma^{\alpha}\overrightarrow{\nabla}_j+\frac{i}{2}\overleftarrow{\nabla}_{j}\gamma^{\alpha}e_{\phantom{.}\alpha}^{j}+m\Big]\psi.
\ee

In curved spacetime, the energy current is related to the energy-momentum tensor via $j_E^{i}=\sqrt{g}\tau^{i}_{\phantom{.}\nu=0}=\sqrt{g}g_{0\mu}\tau^{i\mu }$~\cite{bradlyn2015low}. Because of the Lorentz invariance of massive Dirac fermions, $\tau^{\mu \nu}=\tau^{\nu\mu}$. Thus, we can rewrite $j_E^{i}=\sqrt{g}g_{0\mu}\tau^{\mu i} = \sqrt{g}g_{0\mu}g^{i\nu}\tau_{\phantom{.}\nu}^{\mu}$. Assuming a perturbation of the Luttinger's type~\cite{luttinger1964theory}, the vielbein becomes~\cite{bradlyn2015low} $
e^{\phantom{.}\hat{0}}_{\mu}=\delta^{\phantom{.}\hat{0}}_{\mu}(1+\phi_g)$, and $e^{\phantom{.}a}_{\mu}=\delta^{\phantom{.}a}_{\mu}$. In this case, the expectation value of the energy current is given by~\cite{supplemental_material}
\be \label{eq:J_E_final}
j_E^2(x)=-(1+\phi_g(x))^2\frac{\delta F}{\delta e^{\phantom{.}\hat{2}}_{0}(x)}.
\ee
The partition function and the free energy are defined according to the usual relations as $Z = \text{Tr}\left(e^{-\beta H}\right)$ and $F = -\beta^{-1} \ln Z$, respectively. Here, $\text{Tr}(\ldots)$ denotes the trace in the Fock space~\cite{stoof2009ultracold}.  To derive these equations we used that $\delta e^{0}_{\phantom{.}\alpha}/\delta h^{\phantom{.}\hat{2}}_0=0$ and $\delta \sqrt{g}/\delta h^{\phantom{.}\hat{2}}_0=0$, which we rigorously demonstrate to hold true for a perturbation of the Luttinger's type in~\cite{supplemental_material}.

{\it Boundary fermions}--- Consider a boundary at $x^1=0$ between the gapped bulk at $x^1<0$ with negative mass and that at $x^1>0$ with positive mass (see also Fig.~\ref{fig:one}). The boundary is extended in the whole $x^2$ direction.
The Hamiltonian for boundary fermions can be derived by employing the standard method used in~\cite{nakai2016finite}.  Details are given in~\cite{supplemental_material} for completeness. The full Hamiltonian of Eq.~(\ref{eq:local_Hamiltonian}) is split it into three parts, $H=\int d^2x \psi^{\dagger}[\mathcal{H}_0+\mathcal{H}_1+\mathcal{H}_2]\psi$, where
\be \label{eq:deviation}
\mathcal{H}_0&=&-\frac{i}{2}\gamma^{\hat{0}}\delta^{j}_{\phantom{.}\alpha}\gamma^{\alpha}\overrightarrow{\partial}_j+\frac{i}{2}\overleftarrow{\partial}_j\gamma^{\hat{0}}
\gamma^{\alpha}\delta^j_{\phantom{.}\alpha}
+m,
\nn
\mathcal{H}_1&=&\frac{h}{2}\mathcal{H}_0+(1+\frac{h}{2})(h^{j}_{\phantom{.}\alpha}/2)(i\gamma^{\hat{0}}\gamma^{\alpha}\overrightarrow{\partial}_j-i\overleftarrow{\partial}_j\gamma^{\hat{0}}\gamma^{\alpha}
),
\nn
\mathcal{H}_2&=&(1+\frac{h}{2})\frac{i}{2}(\delta^{\mu}_{\phantom{.}\alpha}-h^{\mu}_{\phantom{.}\alpha}/2)\gamma^{\hat{0}}(\gamma^{\alpha}\omega_{\mu}+\omega_\mu\gamma^{\alpha}).
\ee
Additionally, we defined  $e^{\phantom{.}\alpha}_{\mu}=\delta^{\phantom{.}\alpha}_{\mu}+h^{\phantom{.}\alpha}_{\mu}/2$, $e_{\phantom{.}\alpha}^{\mu}=\delta_{\phantom{.}\alpha}^{\mu}-h_{\phantom{.}\alpha}^{\mu}/2$ and $\sqrt{g}=1+h/2$. We assume the deviations $h$, $h^{\phantom{.}\alpha}_{\mu}$ and $h^{\mu}_{\phantom{.}\alpha}$ in Eq.~(\ref{eq:deviation}) to be small. Therefore, in the following calculations we treat $\mathcal{H}_0$ as the unperturbed Hamiltonian, while $\mathcal{H}_1$ and $\mathcal{H}_2$ are treated perturbatively.
We assume that the metric depends only on $x^2$ near the boundary. Thus, in Luttinger's case~\cite{luttinger1964theory}, $\phi_g(x) = \phi_g(x^2)$. Then the two directions $x^1$ and $x^2$ are completely decoupled in the boundary Hamiltonian. The wave function of the boundary mode obtained from the Hamiltonian $\mathcal{H}_0$ then factorizes into the product of a plane wave in the $x^2$ direction, $\psi_2(x^2)$, and of a two-components evanescent spinor wave function in the $x^1$ direction, $\psi_1(x^1)$: $\psi(x^1,x^2)=\psi_2(x^2)\psi_1(x^1)$. The formal solution of the evanescent spinor is given by~\cite{nakai2016finite}:
\be
\psi_1(x^1)=\exp\left[i\gamma^1\int_0^{x^1}dx'^{1}m(x'^1)\right]\vert s\rangle.
\ee
The two-component spinor $\vert s\rangle$ corresponding to the boundary state satisfies $i\gamma^1\vert s\rangle =\text{sgn}(m)\vert s\rangle$, where $\text{sgn}(m)$ indicates the sign of the mass in the half-space $x^1<0$. The other eigenstate of $i\gamma^1$ corresponds to a state that cannot be normalized~\cite{nakai2016finite}. Therefore, the boundary Hamiltonian obtained from the unperturbed bulk Hamiltonian $\mathcal{H}_0$ is $\tilde{\mathcal{H}}_0=i\text{sgn}(m)\partial_2$.

The derivation of the interaction terms $\mathcal{H}_1$ and $\mathcal{H}_2$ term is more involved. We therefore relegate it to the Supplemental Material~\cite{supplemental_material} and quote only the final result, {\it i.e.} $\tilde{\mathcal{H}}_1=\zeta(x^2)(-\frac{i}{2}(\overrightarrow{\partial}_2-\overleftarrow{\partial}_2))$ and $\tilde{\mathcal{H}}_2=0$, with
\be \label{eq:zeta_def}
\zeta(x^2)
&=&\frac{1}{2}\left(h-h^{2}_{\phantom{.}\hat{2}}-h^{2}_{\phantom{.}\hat{0}}\right)-\frac{1}{4}h\left(h^{2}_{\phantom{.}\hat{2}}+h^{2}_{\phantom{.}\hat{0}}\right),
\ee
where, according to our choice of mass signs, $\sgn(m)=-1$, {\it i.e.} the same as in~\cite{nakai2016finite}. 

We are now in the position to derive the effective boundary free energy at finite temperature, and from it the boundary energy current. To do so, we use the Hamiltonian~(\ref{eq:deviation}) to write the partition function as $Z=\int \mathcal{D}\psi^{*}\mathcal{D}\psi\exp\left(-S^{\rm bdry}[\psi^{*},\psi,\zeta]
\right)$ with imaginary time boundary action~\cite{stoof2009ultracold}
\be
S^{\rm bdry}
= \int_{x^2,\tau}\psi^{*}(x^2,\tau)({\partial_\tau}+\tilde{\mathcal{H}}_0+\tilde{\mathcal{H}}_1)\psi(x^2,\tau),
\ee
where $\int_{x,\tau}=\int^{\beta}_0d\tau\int dx$. Performing the integration over the fermionic fields, the effective free energy functional of the gravitational field is obtained as $F^{\rm bdry}[\zeta]=\beta^{-1}S^{\rm bdry}[\zeta]$. The effective action can be expressed as $S^{\rm bdry}[\zeta]=\sum_{l=1}^{\infty}\text{Tr}[(G_0\Sigma)^l]/l$, where the trace is to be taken over real space $x^2$ and imaginary time $\tau$~\cite{supplemental_material}, up to a constant which is independent of $\zeta$. The inverse  Green's function and self-energy in momentum space are defined as $G_0^{-1}(k,\tau;k',\tau')=-\delta_{k,k'}\left(\partial_\tau+k
\right)\delta(\tau,\tau')$ and $\Sigma(k,\tau;k',\tau')=\big[\zeta(k-k')(k+k')/2\big]\delta(\tau,\tau')$.

At low temperature, the Fermi distribution function $f(p)$ can be approximated using the Sommerfeld expansion as $f(p)\simeq \theta(-p)-(\pi^2 T^2/6)d\delta(p)/dp$. After some lengthy algebra~\cite{supplemental_material}, we obtain the following complete expression for the boundary free energy up to order $T^2$ in the long-wavelength limit~\footnote{Since we neglected temperature-independent terms in our derivation, this form of the boundary free energy is only valid up to a constant which is independent of temperature.}
\be \label{eq:free_energy_result}
F^{\rm bdry}[\zeta]&=&\frac{\pi T^2}{12}\int_{-\infty}^{\infty}dx^2\frac{\zeta(x^2)}{1+\zeta(x^2)}.
\ee
This equation is one of the central results of our paper. In what follows we will use it to derive the boundary energy current and show that it is independent of $\phi_g$ under the long-wavelength approximation.

{\it Energy current}---We begin by recalling Eq.~(\ref{eq:J_E_final}) which we now specify for the boundary case $j^{\rm bdry}_E(x^2) =-
2\big[1+\phi_g(x^2)\big]^2(\delta F^{\rm bdry}[\zeta]/\delta h^{\phantom{.}\hat{2}}_{0}(x^2))$.
Therefore, the energy current flowing along the boundary can be read off from the boundary effective free energy in Eq.~(\ref{eq:free_energy_result}) as
\be \label{eq:boundary_current}
j^{\text{bdry}}_E(x^2)&=&
-2\frac{\pi T^2}{12}\frac{(1+\phi_g(x^2))^2}{(1+\zeta(x^2))^2}\frac{\delta \zeta(x^2)}{\delta h^{\phantom{.}\hat{2}}_{0}(x^2)}.
\ee
The derivation of the functional derivative of $\zeta(x^2)$ is in general a difficult task. In~\cite{supplemental_material} we have carried it out for the case in which $\zeta(x^2)$ is Luttinger's gravitational potential~\cite{luttinger1964theory}, {\it i.e.} a local dilation or contraction of space. Setting $\zeta(x^2)=\phi_g(x^2)$ and we have found that
\be
\label{eq:zeta_function}
\frac{\delta\zeta(x^2)}{\delta h^{\phantom{.}\hat{2}}_{0}(x^2)}\Bigg|_{e^{\phantom{.}\hat{0}}_{\mu}=\delta^{\phantom{.}\hat{0}}_{\mu}(1+\phi_g),
e^{\phantom{.}a}_{\mu}=\delta^{\phantom{.}a}_{\mu}}&=&-\frac{1}{2}.
\ee
Combining Eqs.~(\ref{eq:boundary_current}) and~(\ref{eq:zeta_function}), we obtain the energy boundary current under the long-wavelength approximation given previously in Eq.~(\ref{eq:boundary_current_result}).

Finally, we can use Eq.~(\ref{eq:boundary_current_result}) to calculate the system's thermal Hall conductivity based on the Streda formula~\cite{nomura2012cross,nakai2016finite,zhang2020thermodynamics}. Because of the definition of energy magnetization $M_E^z$ in terms of energy current~\cite{cooper1997thermoelectric}, the boundary energy current satisfies the relation $j^{\text{bdry}}_E=-\big[ M_E^z(x^1=+\infty)-M^z_E(x^1=-\infty)\big]$~\cite{nakai2016finite}, we get $M^z_E=-\sgn(m)\pi T^2/24$. Here, we restored the sign of the mass $m$ using the fact that the bulk energy magnetization $M_E^z$ is odd under parity transformation, so it has opposite signs in the two half-planes: $M_E^z(x^1=-\infty)=-M_E^z(x^1=+\infty)$~\cite{nakai2016finite}. Therefore, the thermal Hall conductivity is then given by the thermal generalization of the Streda formula~\cite{nomura2012cross,nakai2016finite,zhang2020thermodynamics} for the quantized thermal Hall effect
\be
\kappa_H=-\frac{\partial M_E^z}{\partial T}=\sgn(m)\frac{\pi T}{12}.
\ee
This corresponds to a quantized thermal Hall conductivity with Chern number $C=\sgn(m)/2$.

{\it Conclusion}---

In this paper, we consider the boundary modes existing at a domain wall between two (2+1) dimensional massive Dirac fermion systems of opposite masses~\cite{nakai2016finite}. By systematically resumming all-order contributions in powers of the metric tensor at low temperature and in the long-wavelength limit, we have obtained a rigorous expression for the boundary free energy. From this, we have derived the boundary current generated by a gravitational potential of the Luttinger's type (\textit{i.e.} a local dilation or contraction of space). We find that, at least in the low-temperature region, higher-order corrections significantly alter the results of existing first-order calculations~\cite{nakai2016finite}. We show that there is no bulk thermal Hall current proportional to the first derivative of the gravitational potential (\textit{i.e.} proportional to the temperature gradient). Only the boundary supports such contributions, in agreement with numerical simulations~\cite{vinkler2019bulk}. In other words, tidal forces (higher-order gradients) are necessary in order to induce bulk thermal Hall currents~\cite{stone2012gravitational}. Finally, using the generalization of the Streda formula to the thermal Hall effect, we recover the quantized thermal Hall conductivity for (2+1) dimensional massive Dirac fermions with Chern number equal to $\sgn(m)/2$.

{\it Acknowledgments}---
The authors would like to thank Alexander E. Kazantsev for useful discussions. A.P. acknowledges support from the European Commission under the EU Horizon 2020 MSCA-RISE-2019 programme (project 873028 HYDROTRONICS) and from the Leverhulme Trust under the grant agreements RPG-2019-363 and RPG-2023-253. A.D.D.-I. acknowledges support from the Engineering and Physical Sciences Research Council, Grant No. EP/T517823/1.

\onecolumngrid

\balance
\onecolumngrid

\newpage
\clearpage

\vspace{1cm}
\begin{center}
\textbf{\large Supplemental Material: No bulk thermal currents in massive Dirac fermions}
\end{center}

\setcounter{secnumdepth}{3}
\setlength\parindent{0pt}

\setcounter{equation}{0}
\setcounter{figure}{0}
\setcounter{table}{0}
\setcounter{page}{1}

\makeatletter
\renewcommand{\theequation}{S\arabic{equation}}
\renewcommand{\thefigure}{S\arabic{figure}}
\renewcommand{\thetable}{S\arabic{table}}
\renewcommand{\thesection}{S\arabic{section}}

\section{Energy current with finite temperature}
We consider the following model action for a $(2+1)$-dimensional Dirac fermion field coupled with a gravitational field
\be \label{eq:action_def_s}
S&=&\int dtd^2x \sqrt{g} \bar\psi \left[
\frac{i}{2}(e_{\phantom{.}\alpha}^{\mu}\gamma^{\alpha} 
\overrightarrow{\nabla}_{\mu} - \overleftarrow{\nabla}_{\mu}\gamma^{\alpha}e_{\phantom{.}\alpha}^{\mu})-m
\right]\psi
\nn
&=&\int dtd^2x\sqrt{g}\bar{\psi}\left[\frac{i}{2}e^{0}_{\phantom{.}\alpha}\gamma^{\alpha}\overrightarrow{\nabla}_0+\frac{i}{2}e^{j}_{\phantom{.}\alpha}\gamma^{\alpha}\overrightarrow{\nabla}_j-\frac{i}{2}\overleftarrow{\nabla}_{0}\gamma^{\alpha}e_{\phantom{.}\alpha}^{0}-\frac{i}{2}\overleftarrow{\nabla}_{j}\gamma^{\alpha}e_{\phantom{.}\alpha}^{j}
-m\right]\psi
\nn
&=&\int dtd^2x\sqrt{g}\bar{\psi}\left[\frac{i}{2}e^{0}_{\phantom{.}\alpha}\gamma^{\alpha}(\overrightarrow{\partial}_0-\omega_0)-\frac{i}{2}(\overleftarrow{\partial}_{0}+\omega_0)\gamma^{\alpha}e_{\phantom{.}\alpha}^{0}
+\frac{i}{2}e^{j}_{\phantom{.}\alpha}\gamma^{\alpha}\overrightarrow{\nabla}_j-\frac{i}{2}\overleftarrow{\nabla}_{j}\gamma^{\alpha}e_{\phantom{.}\alpha}^{j}
-m\right]\psi
\nn
&=&\int dtd^2x\sqrt{g}\bar\psi\left[\frac{i}{2}e^{0}_{\phantom{.}\alpha}\gamma^{\alpha}\overrightarrow{\partial}_0-\frac{i}{2}\overleftarrow{\partial}_{0}\gamma^{\alpha}e_{\phantom{.}\alpha}^{0}
\right]\psi-\int dt H.
\ee
Here, the covariant derivative $\overrightarrow{\nabla}_{\mu}$ ($\overleftarrow{\nabla}_{\mu}$) acts on the right (left) two-component spinor field $\psi$. Explicitly,
\be
\overrightarrow{\nabla}_{\mu}\psi=\left(\overrightarrow{\partial}_{\mu}+\frac{1}{8}[\gamma_{\alpha},\gamma_\beta]\omega_{\mu}^{\phantom{.}\alpha\beta}
\right)\psi,
\qquad
\bar\psi\overleftarrow{\nabla}_{\mu}=\bar{\psi}\left(\overleftarrow{\partial}_{\mu}-\frac{1}{8}[\gamma_{\alpha},\gamma_\beta]\omega_{\mu}^{\phantom{.}\alpha\beta}
\right).
\ee
From Eq.~(\ref{eq:action_def_s}), we define the energy-momentum tensor
\be
\tau^{\mu}_{\phantom{.}\nu}=e^{\phantom{.}\alpha}_{\nu}\tau^{\mu}_{\phantom{.}\alpha}=
-e^{\phantom{.}\alpha}_{\nu}\frac{1}{\sqrt{g}}\frac{\delta S}{\delta e_{\mu}^{\phantom{.}\alpha}},
\ee
and the Hamiltonian acting in an effectively locally-flat space-time as
\be \label{eq:full_Ham_s}
H&=&\int d^2x \sqrt{g}\bar\psi\left[\frac{i}{2}e^{0}_{\phantom{.}\alpha}\gamma^\alpha\omega_0+\frac{i}{2}\omega_0\gamma^\alpha e^{0}_{\phantom{.}\alpha}
-\frac{i}{2}e^{j}_{\phantom{.}\alpha}\gamma^{\alpha}\overrightarrow{\nabla}_j+\frac{i}{2}\overleftarrow{\nabla}_{j}\gamma^{\alpha}e_{\phantom{.}\alpha}^{j}+m\right]\psi
\nn
&=&\int d^2x\sqrt{g}\psi^{\dagger}\left[\frac{i}{2}e^{0}_{\phantom{.}\alpha}\gamma^{\hat{0}}\gamma^\alpha\omega_0+\frac{i}{2}\gamma^{\hat{0}}\omega_0\gamma^\alpha e^{0}_{\phantom{.}\alpha}
-\frac{i}{2}e^{j}_{\phantom{.}\alpha}\gamma^{\hat{0}}\gamma^{\alpha}\overrightarrow{\nabla}_j+\frac{i}{2}\gamma^{\hat{0}}\overleftarrow{\nabla}_{j}\gamma^{\alpha}e_{\phantom{.}\alpha}^{j}+m\gamma^{\hat{0}}\right]\psi.
\ee
In curved spacetime, the energy current and the energy-momentum tensor satisfy the following relationship
\be
j_E^{i}&=&\sqrt{g}\tau^{i}_{\phantom{.}\nu=0}
=\sqrt{g}g_{0\mu}\tau^{i\mu }.
\ee
Because of the Lorentz invariance of massive Dirac fermions, $\tau^{\mu \nu}=\tau^{\nu\mu}$. Thus, we can rewrite
\be
j_E^{i}=\sqrt{g}g_{0\mu}\tau^{\mu i}
= \sqrt{g}g_{0\mu}g^{i\nu}\tau_{\phantom{.}\nu}^{\mu}=-g_{0\mu}g^{i\nu}e^{\phantom{.}\alpha}_{\nu}\frac{\delta S}{\delta e_{\mu}^{\phantom{.}\alpha}}.
\ee
Assuming a perturbation of the Luttinger’s type~\cite{bradlyn2015low}, the vielbein field becomes
\be \label{eq:luttinger_s}
e^{\phantom{.}\hat{0}}_{\mu}=\delta^{\phantom{.}\hat{0}}_{\mu}(1+\phi_g),
\quad
e^{\phantom{.}a}_{\mu}=\delta^{\phantom{.}a}_{\mu},
\ee
and therefore the energy current is given by
\be \label{eq:J_E_final_s}
j^2_E&=&-\sqrt{g}g_{00}\tau^{0}_{\phantom{.}2}
=
(g_{00})e^{\phantom{.}\alpha}_{2}\frac{\delta S}{\delta e^{\phantom{.}\alpha}_{0}}=
2(1+\phi_g)^2\frac{\delta S}{\delta h^{\phantom{.}\hat{2}}_{0}}.
\ee
In these expressions, $h$, $h^{\phantom{.}\alpha}_{\mu}$ and $h^{\mu}_{\phantom{.}\alpha}$ are treated as generic deviations of the vielbein field and its determinant from their value in a flat space-time:
\be \label{eq:deviation_s}
e^{\phantom{.}\alpha}_{\mu}&=&\delta^{\phantom{.}\alpha}_{\mu}+h^{\phantom{.}\alpha}_{\mu}/2,
\nn
e_{\phantom{.}\alpha}^{\mu}&=&\delta_{\phantom{.}\alpha}^{\mu}-h_{\phantom{.}\alpha}^{\mu}/2,
\nn
\sqrt{g}&\equiv&\det(e^{\phantom{.}\alpha}_{\mu})=1+h/2.
\ee
Now, by substituting the specific expression of the action into Eq.~(\ref{eq:J_E_final_s}), we obtain
\be
\frac{\delta S}{\delta h^{\phantom{.}\hat{2}}_{0}(x,t)}&=&\frac{\delta }{\delta h^{\phantom{.}\hat{2}}_{0}(x,t)}\int dtd^2x\sqrt{g}\bar\psi\left[\frac{i}{2}e^{0}_{\phantom{.}\alpha}\gamma^{\alpha}\overrightarrow{\partial}_0-\frac{i}{2}\overleftarrow{\partial}_{0}\gamma^{\alpha}e_{\phantom{.}\alpha}^{0}
\right]\psi-\frac{\delta }{\delta h^{\phantom{.}\hat{2}}_{0}(x,t)}\int dt H
\nn
&=&0-\frac{\delta }{\delta h^{\phantom{.}\hat{2}}_{0}(x,t)}\int dt H=-\frac{\delta H}{\delta h^{\phantom{.}\hat{2}}_{0}(x)}.
\ee
Here, we employed the relations $\delta e^{0}_{\phantom{.}\alpha}/\delta h^{\phantom{.}\hat{2}}_0=0$ and $\delta \sqrt{g}/\delta h^{\phantom{.}\hat{2}}_0=0$, which we rigorously demonstrate to hold true for a Luttinger's gravitational potential in Section~\ref{sect:Zeta_Fun}. Besides, it should be noted that the final step of the above derivation holds only in the case of a static vielbein field (independent of time), which is precisely the scenario we are considering. Therefore, the expression of the energy current  Eq.~(\ref{eq:J_E_final_s}) becomes
\be
j_E^2(x)=-2(1+\phi_g)^2\frac{\delta H}{\delta h^{\phantom{.}\hat{2}}_{0}(x)}=-(1+\phi_g)^2\frac{\delta H}{\delta e^{\phantom{.}\hat{2}}_{0}(x)}.
\ee
Thus, at finite temperature $T$, the expectation value of the energy current is given by
\be \label{eq:average_action_derivative_s}
j_E^2(x)&=&-2(1+\phi_g(x))^2\frac{1}{Z}\text{Tr}\left(e^{-\beta H} \frac{\delta H}{\delta h^{\phantom{.}\hat{2}}_{0}(x)}\right)
=2\frac{(1+\phi_g(x))^2}{Z\beta} \frac{\delta Z}{\delta h^{\phantom{.}\hat{2}}_{0}(x)} 
\nn
&=&2\frac{(1+\phi_g(x))^2}{Z\beta} \frac{\delta e^{-\beta F}}{\delta h^{\phantom{.}\hat{2}}_{0}(x)}
=-2(1+\phi_g(x))^2\frac{\delta F}{\delta h^{\phantom{.}\hat{2}}_{0}(x)}=-(1+\phi_g(x))^2\frac{\delta F}{\delta e^{\phantom{.}\hat{2}}_{0}(x)}
.
\ee
Here, we have introduced the partition function
\be \label{eq:part_function_s}
&&Z = \text{Tr}\left(e^{-\beta H}\right)=\int \mathcal{D}[\psi^*,\psi]\exp\left(-S^E[\psi^*,\psi]
\right)
,
\nn
&&S^{E}[\psi^*,\psi]=\int_0^{\beta}d\tau\Big(\int d^2x\psi^{*}(x,\tau)\frac{\partial}{\partial \tau}\psi(x,\tau)+H[\psi^*(\tau),\psi(\tau)]\Big),
\ee
and the free energy, according to the usual relation $F = -\beta^{-1} \ln Z$.  
Equations~(\ref{eq:average_action_derivative_s}) and~(\ref{eq:part_function_s}) are the fundamental results of this section, {\it i.e.} are the expressions that we will use in what follows to calculate the energy current.
We stress that these formulae remain valid for fermions living at an edge or a boundary between regions of space with different physical properties (e.g., different masses). Since we are interested in the energy current carried by fermions localised at a boundary, in what follows we will first derive the Hamiltonian and free energy of such particles.

\section{Boundary Hamiltonian}
In this section, we establish the boundary Hamiltonian using the same method as in Ref.~\cite{nakai2016finite}. Using the definitions introduced in Eq.~(\ref{eq:deviation_s}), the Hamiltonian of Eq.~(\ref{eq:full_Ham_s}) is split into three parts
\be
\mathcal{H}_0&=&-\frac{i}{2}\gamma^{\hat{0}}\delta^{j}_{\phantom{.}\alpha}\gamma^{\alpha}\overrightarrow{\partial}_j+\frac{i}{2}\overleftarrow{\partial}_j\gamma^{\hat{0}}
\gamma^{\alpha}\delta^j_{\phantom{.}\alpha}
+m,
\nn
\mathcal{H}_1&=&\frac{h}{2}\mathcal{H}_0+\left(1+\frac{h}{2}\right)(h^{j}_{\phantom{.}\alpha}/2)(i\gamma^{\hat{0}}\gamma^{\alpha}\overrightarrow{\partial}_j-i\overleftarrow{\partial}_j\gamma^{\hat{0}}\gamma^{\alpha}
),
\nn
\mathcal{H}_2&=&\left(1+\frac{h}{2}\right)\left[\frac{i}{2}\delta^{\mu}_{\phantom{.}\alpha}\gamma^{\hat{0}}(\gamma^{\alpha}\omega_{\mu}+\omega_\mu\gamma^{\alpha})-\frac{i}{2}(h^{\mu}_{\phantom{.}\alpha}/2)\gamma^{\hat{0}}(\gamma^{\alpha}\omega_{\mu}+\omega_\mu\gamma^{\alpha})\right].
\ee
We consider a boundary at $x^1=0$ between a gapped bulk of mass $m<0$ (\textit{i.e.} in the half-plane $x^1<0$) and mass $m>0$ (\textit{i.e.} in the half-plane $x^1>0$). The boundary is extended to the entire $x^2$ space.
This derivation is based on an assumption that the metric depends only on $x^2$ near the boundary. Then, $x^1$ and $x^2$ are completely decoupled in the boundary Hamiltonian. The wave function of the boundary mode of the Hamiltonian $\mathcal{H}_0$ is a product of a plane wave of the $x^2$-coordinate and a two-components spinor wave function of the $x^1$-coordinate
\be
\psi(x^1,x^2)=\psi_2(x^2)\psi_1(x^1).
\ee
The boundary modes satisfy following equation~\cite{nakai2016finite}
\be \label{eq:free_Dirac_s}
&&(-i\gamma^{\hat{0}}\gamma^{\hat{1}}\partial_1+m(x^1)\gamma^{\hat{0}})\psi_1(x^1)=0,
\nn
&&-i\partial_1\psi_1^{\dagger}(x^1)\gamma^{\hat{0}}\gamma^{\hat{1}}-m(x^1)\psi_1^{\dagger}(x^1)\gamma^{\hat{0}}=0.
\ee
Formally, the solution of (\ref{eq:free_Dirac_s}) is given by 
\be
\psi_1(x^1)=\exp\left[i\gamma^{\hat{1}}\int_0^{x^1}dx'^{1}m(x'^1)\right]\vert s\rangle.
\ee
Here, the two component spinor $\vert s\rangle$ corresponding to the edge bound states satisfies $i\gamma^1\vert s\rangle =\text{sgn}(m)\vert s\rangle $, with $\text{sgn}(m)$ indicating the sign of the mass in the half-plane $x^1<0$. The other one corresponds to states that cannot be normalized~\cite{nakai2016finite}. The boundary Hamiltonian for the unperturbed bulk Hamiltonian $\mathcal{H}_0$ is
\be
\tilde{\mathcal{H}}_0&=&\langle \psi_1\vert \mathcal{H}_0\vert\psi_1\rangle
=\langle \psi_1\vert(-\frac{i}{2}\gamma^{\hat{0}}\gamma^{\hat{1}}\overrightarrow{\partial}_1+
\frac{i}{2}\overleftarrow{\partial}_1\gamma^{\hat{0}}\gamma^{\hat{1}}-
\frac{i}{2}\gamma^{\hat{0}}\gamma^{\hat{2}}\overrightarrow{\partial}_2
+\frac{i}{2}\overleftarrow{\partial}_2\gamma^{\hat{0}}\gamma^{\hat{2}}
+m\gamma^{\hat{0}})\vert\psi_1\rangle
\nn
&=&\langle s\vert -\frac{i}{2}\gamma^{\hat{0}}\gamma^{\hat{2}}\overrightarrow{\partial}_2
+\frac{i}{2}\overleftarrow{\partial}_2\gamma^{\hat{0}}\gamma^{\hat{2}}
\vert s\rangle =\frac{i}{2}\text{sgn}(m)\overrightarrow{\partial}_2-\frac{i}{2}\sgn(m)\overleftarrow{\partial}_2,
\ee
where we have used the relation $i\gamma^{\hat{1}}=-\gamma^{\hat{0}}\gamma^{\hat{2}}$. The derivation of the interaction terms $\mathcal{H}_1$ and $\mathcal{H}_2$ term is more involved. We start by defining their corresponding boundary terms:
\be
\tilde{\mathcal{H}}_1&=&\frac{h}{2}\langle \psi_1\vert \mathcal{H}_0\vert \psi_1\rangle+\left(1+\frac{h}{2}\right)(h^{j}_{\phantom{.}\alpha}/2)\langle \psi_1\vert \frac{i}{2}\gamma^{\hat{0}}\gamma^{\alpha}\overrightarrow{\partial}_j -\frac{i}{2}\overleftarrow{\partial}_j\gamma^{\hat{0}}\gamma^{\alpha}
\vert \psi_1\rangle,
\nn
\tilde{\mathcal{H}}_2&=&\left(1+\frac{h}{2}\right)\frac{1}{2}\Big(\langle \psi_1\vert i\gamma^{\hat{0}}\delta^{\mu}_{\phantom{.}\alpha}(\gamma^{\alpha}\omega_{\mu}+\omega_{\mu}\gamma^{\alpha})\vert \psi_1\rangle-\langle \psi_1\vert i\gamma^{\hat{0}}(h^{\mu}_{\phantom{.}\alpha}/2)(\gamma^{\alpha}\omega_{\mu}+\omega_{\mu}\gamma^{\alpha})\vert \psi_1\rangle\Big).
\ee
In order to project $\mathcal{H}_1$ and $\mathcal{H}_2$ onto the boundary mode, we need to consider the following three terms:
\be
A&=&(h^{j}_{\phantom{.}\alpha}/2)\langle \psi_1\vert \frac{i}{2}\gamma^{\hat{0}}\gamma^{\alpha}\overrightarrow{\partial}_j-\frac{i}{2}\overleftarrow{\partial}_j\gamma^{\hat{0}}\gamma^{\alpha}
\vert\psi_1\rangle,
\nn
B&=&B_1+B_2=
i\omega_{\mu xy}\langle \psi_1\vert \gamma^{\hat{0}}\gamma^{\mu}\gamma^x\gamma^y \vert \psi_1\rangle+i\omega_{\mu xy}\langle \psi_1\vert \gamma^{\hat{0}}\gamma^x\gamma^y \gamma^{\mu}\vert \psi_1\rangle,
\nn
C&=&C_1+C_2=i\omega_{\mu xy}(h_{\phantom{.}\alpha}^{\mu}/2)\langle \psi_1\vert \gamma^{\hat{0}}\gamma^{\alpha}\gamma^x\gamma^y \vert \psi_1\rangle+i\omega_{\mu xy}(h_{\phantom{.}\alpha}^{\mu}/2)\langle \psi_1\vert \gamma^{\hat{0}}\gamma^x\gamma^y \gamma^{\alpha}\vert \psi_1\rangle.
\ee
Notice that $h^{\phantom{.}j}_{\alpha}$, $h$, and $\omega_{\mu \alpha\beta}$ are just functions of $x^2$, and $\omega_{\mu \alpha\beta}=-\omega_{\mu \beta\alpha}$. A is given by
\be \label{eq:term_A_s}
A&=&
(h^{1}_{\phantom{.}\alpha}/2)\langle \psi_1\vert \frac{i}{2}\gamma^{\hat{0}}\gamma^{\alpha}\overrightarrow{\partial}_1-\frac{i}{2}\overleftarrow{\partial}_1\gamma^{\hat{0}}\gamma^{\alpha}\vert\psi_1\rangle
+(h^{2}_{\phantom{.}\alpha}/2)\langle \psi_1\vert \frac{i}{2}\gamma^{\hat{0}}\gamma^{\alpha}\overrightarrow{\partial}_2-\frac{i}{2}\overleftarrow{\partial}_2\gamma^{\hat{0}}\gamma^{\alpha}\vert\psi_1\rangle
\nn
&
=&\frac{1}{2}(h_{\phantom{.}\alpha}^{1}(x^2)/2)\langle \psi_1\vert -\gamma^{\hat{0}}\gamma^\alpha\gamma^{\hat{1}}m(x^1)+m(x_1)\gamma^{\hat{1}}\gamma^{\hat{0}}\gamma^{\alpha}
\vert\psi_1\rangle 
+(h_{\phantom{.}\alpha}^{2}(x^2)/2)\langle s\vert \gamma^{\hat{0}}\gamma^{\alpha}\vert s\rangle \frac{i}{2}(\overrightarrow{\partial}_2-\overleftarrow{\partial}_2).
\ee
The first term in Eq.~(\ref{eq:term_A_s}) is zero since it contains following integral~\cite{nakai2016finite} (the integrand of which is an odd function)
\be
\int_{-\infty}^{\infty}dx^1m(x^1)\exp\left[2\text{sgn}(m)\int_0^{x^1}dx'^1m(x'^1)
\right]=0.
\ee
The second term in Eq.~(\ref{eq:term_A_s}) is nonzero for $\alpha=0,2$ due to the property of the two-component spinor $\vert s\rangle$
\be
\langle s\vert \gamma^{\hat{0}}\gamma^{\hat{0}}\vert s\rangle&=&1,
\nn
\langle s\vert \gamma^{\hat{0}}\gamma^{\hat{1}}\vert s\rangle&=&-i\text{sgn}(m)\langle s\vert \gamma^0\vert s\rangle=0,
\nn
\langle s\vert \gamma^{\hat{0}}\gamma^{\hat{2}}\vert s\rangle&=&-\text{sgn}(m).
\ee
Here we have used $\gamma^{\hat{1}}(\gamma^{\hat{0}}\vert s\rangle) =-\gamma^{\hat{0}}\gamma^{\hat{1}}\vert s\rangle$. Therefore,
\be
A=[(h_{\phantom{.}{\hat{0}}}^{2}(x^2)/2)-\sgn(m)(h_{\phantom{.}{\hat{2}}}^{2}(x^2)/2)]\frac{i}{2}(\overrightarrow{\partial}_2-\overleftarrow{\partial}_2).
\ee
Now we consider the terms B and C. It is readily apparent that the values of these matrix elements are
\be 
&&\langle s\vert \gamma^{\hat{0}}\gamma^{\hat{0}}\gamma^{\hat{0}}\gamma^{\hat{0}}\vert s\rangle=\langle s\vert \gamma^{\hat{0}}\gamma^{\hat{1}}\gamma^{\hat{0}}\gamma^{\hat{1}}\vert s\rangle=\langle s\vert \gamma^{\hat{0}}\gamma^{\hat{2}}\gamma^{\hat{0}}\gamma^{\hat{2}}\vert s\rangle=1,
\nn
&&\langle s\vert \gamma^{\hat{0}}\gamma^{\hat{0}}\gamma^{\hat{1}}\gamma^{\hat{1}}\vert s\rangle=\langle s\vert \gamma^{\hat{0}}\gamma^{\hat{1}}\gamma^{\hat{1}}\gamma^{\hat{0}}\vert s\rangle=\langle s\vert \gamma^{\hat{0}}\gamma^{\hat{0}}\gamma^{\hat{2}}\gamma^{\hat{2}}\vert s\rangle=\langle s\vert \gamma^{\hat{0}}\gamma^{\hat{2}}\gamma^{\hat{2}}\gamma^{\hat{0}}\vert s\rangle=-1,
\nn
&&\langle s\vert \gamma^{\hat{0}}\gamma^{\hat{2}}\gamma^{\hat{2}}\gamma^{\hat{2}}\vert s\rangle=\langle s\vert \gamma^{\hat{0}}\gamma^{\hat{0}}\gamma^{\hat{2}}\gamma^{\hat{0}}\vert s\rangle=\langle s\vert \gamma^{\hat{0}}\gamma^{\hat{1}}\gamma^{\hat{1}}\gamma^{\hat{2}}\vert s\rangle=\langle s\vert \gamma^{\hat{0}}\gamma^{\hat{2}}\gamma^{\hat{1}}\gamma^{\hat{1}}\vert s\rangle=\text{sgn}(m),
\nn
&&\langle s\vert \gamma^{\hat{0}}\gamma^{\hat{0}}\gamma^{\hat{0}}\gamma^{\hat{2}}\vert s\rangle=\langle s\vert \gamma^{\hat{0}}\gamma^{\hat{2}}\gamma^{\hat{0}}\gamma^{\hat{0}}\vert s\rangle=\langle s\vert \gamma^{\hat{0}}\gamma^{\hat{1}}\gamma^{\hat{2}}\gamma^{\hat{1}}\vert s\rangle=-\text{sgn}(m),
\ee
with all other matrix elements being zero. Thus, for terms B and C:
\be
&&B_1=2i(\omega_{1{\hat{0}}{\hat{1}}}-\omega_{2{\hat{2}}{\hat{0}}})+2i\text{sgn}(m)(\omega_{0{\hat{2}}{\hat{0}}}+\omega_{1{\hat{1}}{\hat{2}}}),
\nn
&&B_2=-2i(\omega_{1\hat{0}\hat{1}}-\omega_{2\hat{2}\hat{0}})-2i\sgn(m)(\omega_{0\hat{2}\hat{0}}+\omega_{1\hat{1}\hat{2}}),
\nn
&&B=B_1+B_2=0,
\nn
&&C_1=i\Big(h^{\mu}_{\phantom{.}{\hat{1}}}\omega_{\mu {\hat{0}}{\hat{1}}}+h^{\mu}_{\phantom{.}{\hat{2}}}\omega_{\mu {\hat{0}}{\hat{2}}}
+\text{sgn}(m)[h^{\mu}_{\phantom{.}{\hat{0}}}\omega_{\mu{\hat{2}}{\hat{0}}}+h^{\mu}_{\phantom{.}{\hat{1}}}\omega_{\mu {\hat{1}}{\hat{2}}}]\Big),
\nn
&&C_2=-i\Big(h^{\mu}_{\phantom{.}{\hat{1}}}\omega_{\mu {\hat{0}}{\hat{1}}}+h^{\mu}_{\phantom{.}{\hat{2}}}\omega_{\mu {\hat{0}}{\hat{2}}}
+\text{sgn}(m)[h^{\mu}_{\phantom{.}{\hat{0}}}\omega_{\mu{\hat{2}}{\hat{0}}}+h^{\mu}_{\phantom{.}{\hat{1}}}\omega_{\mu {\hat{1}}{\hat{2}}}]\Big),
\nn
&&C=C_1+C_2=0.
\ee
Based on this information, we can express the interaction terms $\tilde{\mathcal{H}}_1$ and $\tilde{\mathcal{H}}_2$ as
\be
\tilde{\mathcal{H}}_1&=&\frac{h}{2}\langle \psi_1\vert \mathcal{H}_0\vert \psi_1\rangle+\left(1+\frac{h}{2}\right)\langle \psi_1\vert(i\gamma^{\hat{0}}(h^{j}_{\phantom{.}\alpha}/2)\gamma^{\alpha}\partial_j)\vert \psi_1\rangle
\nn
&=&\frac{h(x^2)}{2}(\frac{i}{2}\sgn(m)(\overrightarrow{\partial}_2-\overleftarrow{\partial}_2))+\frac{1}{2}(1+\frac{h(x^2)}{2})(h_{\phantom{.}{\hat{0}}}^{2}(x^2)-\sgn(m)h_{\phantom{.}{\hat{2}}}^{2}(x^2)
)\frac{i}{2}(\overrightarrow{\partial}_2-\overleftarrow{\partial}_2)
\nn
&=&\frac{1}{2}\Big(-\sgn(m)h(x^2)+(1+\frac{h(x^2)}{2})(\sgn(m)h^{2}_{\phantom{.}\hat{2}}(x^2)-h^{2}_{\phantom{.}\hat{0}}(x^2))\Big)(-\frac{i}{2}(\overrightarrow{\partial}_2-\overleftarrow{\partial}_2)),
\nn
\tilde{\mathcal{H}}_2&=&\left(1+\frac{h}{2}\right)\frac{1}{2}\Big(\langle \psi_1\vert i\gamma^{\hat{0}}\delta^{\mu}_{\phantom{.}\alpha}(\gamma^{\alpha}\omega_{\mu}+\omega_{\mu}\gamma^{\alpha})\vert \psi_1\rangle-\langle \psi_1\vert i\gamma^{\hat{0}}(h^{\mu}_{\phantom{.}\alpha}/2)(\gamma^{\alpha}\omega_{\mu}+\omega_{\mu}\gamma^{\alpha})\vert \psi_1\rangle\Big)=0,
\ee
where we have used the property of spin connection coefficient $\omega_{\mu \alpha\beta}=-\omega_{\mu\beta \alpha}$. It is interesting to note that the spin connection part vanishes in this case, which means that boundary fermions are always massless, regardless of the form of the metric. In the main text, we have chosen $\sgn(m)=-1$. In what follows we will use this sign of the mass term. In this case, $\tilde{\mathcal{H}}_1=\zeta(x^2)(-\frac{i}{2}(\overrightarrow{\partial}_2-\overleftarrow{\partial}_2))$, $\tilde{\mathcal{H}}_2=0$, with
\be \label{eq:zeta_def_s}
\zeta(x^2)
&=&\frac{1}{2}\left(h(x^2)-h^{2}_{\phantom{.}\hat{2}}(x^2)-h^{2}_{\phantom{.}\hat{0}}(x^2)\right)-\frac{1}{4}h(x^2)\left(h^{2}_{\phantom{.}\hat{2}}(x^2)+h^{2}_{\phantom{.}\hat{0}}(x^2)\right),
\ee
and we obtain the boundary Hamiltonian
\be \label{eq:hamiltonian_boundary_final_s}
\tilde{H}=\int dx^2\psi_2^{\dagger}(x^2)
\Big(-\frac{i}{2}(\overrightarrow{\partial}_2-\overleftarrow{\partial}_2)+\zeta(x^2)(-\frac{i}{2}(\overrightarrow{\partial}_2-\overleftarrow{\partial}_2))\Big)\psi_2(x^2).
\ee
\section{Boundary free energy}
We now derive the effective boundary free energy at finite temperature. To do so, we use the Hamiltonian in Eq.~(\ref{eq:hamiltonian_boundary_final_s}) to write the partition function as
\be
&&Z=\int \mathcal{D}\psi^{*}\mathcal{D}\psi\exp\left(-S^{\rm bdry}[\psi^{*},\psi,\zeta]
\right)=\int \mathcal{D}\psi^{*}\mathcal{D}\psi\exp\left(-S^{\rm bdry}_0[\psi^*,\psi]-S^{\rm bdry}_{1}[\psi^*,\psi,\zeta]
\right).
\ee
Here,
\be
S^{\rm bdry}_0[\psi^*,\psi]&=&\int_0^{\beta}d\tau \int dx^2 \psi^*(x^2,\tau)\left[\frac{\partial}{\partial\tau}-\frac{i}{2}\overrightarrow{\frac{\partial}{\partial {x^2}}}+\frac{i}{2}\overleftarrow{\frac{\partial}{\partial {x^2}}}
\right]\psi(x^2,\tau),
\nn
S^{\rm bdry}_1[\psi^*,\psi,\zeta]&=&\int^{\beta}_0 d\tau\int dx^2\psi^*(x^2,\tau)\left[-\frac{i}{2}\zeta(x^2)\overrightarrow{\frac{\partial}{\partial {x^2}}}+\frac{i}{2}\overleftarrow{\frac{\partial}{\partial {x^2}}}\zeta(x^2)
\right]\psi(x^2,\tau).
\ee
Performing the integration over the fermionic fields, the effective free energy functional of the gravitational field is obtained as
\be \label{eq:free_energy_s}
S^{\rm bdry}[\zeta]&=&\beta F^{\rm bdry}[\zeta]
=\sum_{l=1}^{\infty}\frac{1}{l}\text{Tr}[(G_0\Sigma)^l],
\ee
where the trace was taken over real space $x^2$ and imaginary time $\tau$. This expression is exact up to a constant that we have neglected, since it does not contribute to the thermal current. In Eq.~(\ref{eq:free_energy_s}), the inverse  Green's function and self-energy in momentum space, $G_0^{-1}$ and $\Sigma$, respectively,
are defined as 
\be \label{eq:G_0_Sigma_def_s}
&&G_0^{-1}(k,\tau;k',\tau')=-\delta_{k,k'}\left(\partial_\tau+k
\right)\delta(\tau,\tau'),
\nn
&&\Sigma(k,\tau;k',\tau')=\zeta(k-k')\frac{k+k'}{2}\delta(\tau,\tau').
\ee
To calculate the free energy of Eq.~(\ref{eq:free_energy_s}), we first consider the terms corresponding to $l=1,2$. For $l=1$, we get
\be \label{eq:first_order_s}
&&\text{Tr}[G_0\Sigma]=\int dx^2 [\zeta(x^2)I^{(1)}],
\nn
&&I^{(1)}=\sum_{\omega_n}\int \frac{dp}{2\pi}\frac{p}{i\omega_n-p}=\beta \int\frac{dp}{2\pi}pf(p),
\ee
where $\zeta(x^2)$ is given in Eq.~(\ref{eq:zeta_def_s}). At low temperatures, the Fermi distribution function $f(p)$ in Eq.~(\ref{eq:first_order_s}) can be approximated using the Sommerfeld expansion as
\be \label{eq:Sommerfeld_exp_s}
f(p)\simeq \theta(-p)-\frac{\pi^2 T^2}{6}\frac{d\delta(p)}{d p}.
\ee
We immediately see that when this expansion is plugged into $I^{(1)}$, it takes the form
\be \label{eq:first_order_result_s}
I^{(1)}&=&\beta \int \frac{dp}{2\pi}p f(p)=\beta \int \frac{dp}{2\pi}p\left(\theta(-p)-\frac{\pi^2 T^2}{6}\frac{d\delta(p)}{d p}
\right)=\frac{\beta}{2\pi}\Lambda+\beta \frac{\pi T^2}{12}.
\ee
Here, $\Lambda$ is a large constant that is independent of temperature. We neglect its effect, since it produces a contribution to the free energy that is independent of temperature.  In fact, the first term on the right-hand side of $I^{(1)}$ [{\it i.e.}, the one proportional to $\theta(-p)$] produces a contribution to the free energy which is independent of temperature, and therefore we can neglect it~\footnote{Note indeed that, according to Eq.~(\ref{eq:free_energy_s}), to obtain $F$ we have to multiply by $\beta^{-1}$}. On the other hand, the second term of Eq.~(\ref{eq:first_order_result_s}) is proportional to the temperature and must be retained. Therefore,
\be
\text{Tr}[G_0\Sigma]=\beta \frac{\pi T^2}{12}\int dx^2 \zeta(x^2).
\ee
For the term with $l=2$ in Eq.~(\ref{eq:free_energy_s}), we can write
\be \label{eq:second_order_s}
\frac{1}{2}\text{Tr}[(G_0\Sigma)^2]
&=&\int \frac{dq}{2\pi} \int dx^2 d{y}^2 e^{iq(x^2-y^2)}
\Pi(q)\zeta(x^2)\zeta(y^2),
\ee
where, for $q=0$,
\be 
\Pi(0)&=&\frac{1}{2}\sum_{\omega_n}\int \frac{dp}{2\pi}
\frac{p^2}{(i\omega_n-p)^2}=\frac{\beta}{2}\int \frac{dp}{2\pi}\frac{df(p)}{dp}p^2.
\ee
We only need to consider $\Pi(q)$ at $q=0$ because, from Eq.~(\ref{eq:second_order_s}), it can be readily seen that if $\Pi(q)$ is expanded in powers of $q$. In that case, the contributions due to higher-order terms are proportional to higher-order derivatives of $\zeta(x^2)$. However, the primary focus here is on contributions to the free energy that do not contain such derivatives, and therefore we will neglect these terms in what follows and set $q=0$ in $\Pi(q)$:
\be
\Pi(0)&=&\frac{\beta}{2}\int \frac{dp}{2\pi}p^2\frac{d}{dp}\left(\theta(-p)-\frac{\pi^2 T^2}{6}\frac{d\delta(p)}{d p}
\right)=-\beta \frac{\pi T^2}{12}.
\ee
Here, we used the Sommerfeld expansion given in Eq.~(\ref{eq:Sommerfeld_exp_s}), thus
\be
\frac{1}{2}\text{Tr}[(G_0\Sigma)^2]
&=&\int \frac{dq}{2\pi} \int dx^2 d{y}^2 e^{iq(x^2-y^2)}\Pi(0)\zeta(x^2)\zeta(y^2)
=\int dx^2 \Pi(0)\zeta(x^2)\zeta(x^2)
=-\beta\frac{\pi T^2}{12}\int dx^2(\zeta(x^2))^2.
\ee

Let us now consider the coefficients of order $l>2$ in Eq.~(\ref{eq:free_energy_s}). Since, as shown above, we only need to consider coefficients at $q = 0$, the $l$-th order term of the expansion can be generally represented as
\be
\frac{1}{l}\text{Tr}[(G_0\Sigma)^l]&=&\int \frac{dq_1\cdots dq_{l-1}}{(2\pi)^{l-1}}\int dx_1^2dx_2^2\cdots dx_l^2U_{l}\exp\left(i\sum_{j=1}^{l-1}q_jx_j^2-i(\sum_{j=1}^{l-1}q_j)x_l^2
\right)\zeta(x_l^2)\zeta(x_l^2)\cdots \zeta(x_l^2)
\nn
&=&\frac{1}{l}U_{l}\int dx^2 (\zeta (x^2))^{l},
\nn
U_{l}&=&\sum_{\omega_n}\int \frac{dk}{2\pi}k^{l}\left(\frac{1}{i\omega_n-k}\right)^l.
\ee
Following Hertz's approach~\cite{hertz1974fluctuations}, we convert the above equation into a contour integral and perform the Matsubara sum, which gives
\be \label{eq:U_nm_s}
U_{l}=-\frac{\beta}{2\pi i}\int \frac{dk}{2\pi}k^{l}\oint_c dz \frac{f(z)}{(z-k)^l}
=\frac{\beta}{(l-1)!} \int \frac{dk}{2\pi}k^{l}f^{(l-1)}(k).
\ee
Here, $f^{(n)}(k)$ denotes the $n$th derivative of the function $f(k)$. Substituting Eq.~(\ref{eq:Sommerfeld_exp_s}) into Eq.~(\ref{eq:U_nm_s}) and using the property of Dirac delta function as mentioned above, it is possible to show that
\be
k^lf^{(l-1)}(k)&=&-k^l\delta^{(l-2)}(k)-\frac{\pi^2 T^2}{6}k^l\delta^{(l)}(k)
=-(-1)^{l-2}[k^l]^{(l-2)}\delta(k)-\frac{\pi^2 T^2}{6}(-1)^l[k^l]^{(l)}\delta(k)
\nn
&=&-(-1)^l\delta(k)\left(\frac{l!}{2}k^2+\frac{\pi^2T^2}{6}l!
\right)=(-1)^{l+1}l!(\frac{k^2}{2}+\frac{\pi^2T^2}{6})\delta(k),
\ee
such that we have 
\be
U_{l}=(-1)^{l+1}\frac{\beta l!}{(l-1)!} \int \frac{dk}{2\pi}(\frac{k^2}{2}+\frac{\pi^2T^2}{6})\delta(k)=(-1)^{l+1}l\beta \frac{\pi T^2}{12}.
\ee
Therefore, the contribution from the $l$th order term is given by
\be
\frac{1}{l}\text{Tr}[(G_0\Sigma)^l]
&=&\frac{1}{l}(-1)^{l+1}l\beta\frac{\pi T^2}{12}
\int dx^2(\zeta(x^2))^l=\beta (-1)^{l+1}\frac{\pi T^2}{12}\int dx^2(\zeta(x^2))^l.
\ee

By summing these contributions together as dictated by Eq.~(\ref{eq:free_energy_s}), we obtain the following complete expression for the boundary free energy
\be
F^{\rm bdry}[\zeta]&=&\frac{\pi T^2}{12}\int_{-\infty}^{\infty}dx^2\frac{\zeta(x^2)}{1+\zeta(x^2)}.
\ee
This equation is one of the central results of our paper.

\section{Function derivative of zeta function} \label{sect:Zeta_Fun}
Due to the property of vielbein field~\cite{carroll2019spacetime}: $e^{\phantom{.}\alpha}_{\mu}e^{\nu}_{\phantom{.}\alpha}=\delta^\nu_{\mu}$, $e^{\phantom{.}\alpha}_{\mu}e^{\mu}_{\phantom{.}\beta}=\delta^{\alpha}_{\beta}$ and Eq.~(\ref{eq:deviation_s}), we define
\be
H_{\mu\alpha}&=&e^{\phantom{.}\alpha}_{\mu}=\left(
\begin{matrix}
1+h^{\phantom{.}\hat{0}}_{0}/2&h^{\phantom{.}\hat{1}}_{0}/2&h^{\phantom{.}\hat{2}}_{0}/2
\\
h^{\phantom{.}\hat{0}}_{1}/2&1+h^{\phantom{.}\hat{1}}_{1}/2&h^{\phantom{.}\hat{2}}_{1}/2
\\
h^{\phantom{.}\hat{0}}_{2}/2&h^{\phantom{.}\hat{1}}_{2}/2&1+h^{\phantom{.}\hat{2}}_{2}/2
\end{matrix}
\right),
\nn
A_{\beta\nu}&=&e^{\nu}_{\phantom{.}\beta}=\left(
\begin{matrix}
1-h^{0}_{\phantom{.}\hat{0}}/2&-h^{1}_{\phantom{.}\hat{0}}/2&-h_{\phantom{.}\hat{0}}^{2}/2
\\
-h_{\phantom{.}\hat{1}}^{0}/2&1-h_{\phantom{.}\hat{1}}^{1}/2&-h_{\phantom{.}\hat{1}}^{2}/2
\\
-h_{\phantom{.}\hat{2}}^{0}/2&-h_{\phantom{.}\hat{2}}^{1}/2&1-h_{\phantom{.}\hat{2}}^{2}/2
\end{matrix}
\right).
\ee
Obviously, $H.A=A.H=1$. Using this, we can express $h^{\mu}_{\phantom{.}\alpha}$ in terms of $h^{\phantom{.}\alpha}_{\mu}$. To calculate the functional derivative of $\zeta(x^2)$, we need the three terms $h$, $h^{2}_{\phantom{.}\hat{0}}$, and $h^{2}_{\phantom{.}\hat{2}}$:
\be \label{eq:h0h2_s}
h^{2}_{\phantom{.}\hat{0}}&=&\frac{1}{2}\Big(2h^{\phantom{.}\hat{2}}_0+h^{\phantom{.}\hat{1}}_1h^{\phantom{.}\hat{2}}_0-h^{\phantom{.}\hat{1}}_0h^{\phantom{.}\hat{2}}_1\Big)\Big(1+\frac{h}{2}\Big)^{-1},
\nn
h^{2}_{\phantom{.}\hat{2}}&=&2-\frac{1}{2}\Big(4+2h^{\phantom{.}\hat{0}}_0-h^{\phantom{.}\hat{0}}_1h^{\phantom{.}\hat{1}}_0+2h^{\phantom{.}\hat{1}}_1+h^{\phantom{.}\hat{0}}_0h^{\phantom{.}\hat{1}}_1\Big)\Big(1+\frac{h}{2}\Big)^{-1},
\nn
h^0_{\phantom{.}\hat{0}}&=&2-\frac{1}{2}\Big(4+2h^{\phantom{.}\hat{1}}_1-h^{\phantom{.}\hat{1}}_2h^{\phantom{.}\hat{2}}_1+2h^{\phantom{.}\hat{2}}_2+h^{\phantom{.}\hat{1}}_1h^{\phantom{.}\hat{2}}_2\Big)\Big(1+\frac{h}{2}\Big)^{-1},
\nn
h^{0}_{\phantom{.}\hat{1}}&=&\frac{1}{2}\Big(2h^{\phantom{.}\hat{0}}_1-h^{\phantom{.}\hat{0}}_2h^{\phantom{.}\hat{2}}_1+h^{\phantom{.}\hat{0}}_1h^{\phantom{.}\hat{2}}_2
\Big)\Big(1+\frac{h}{2}\Big)^{-1},
\nn
h^{0}_{\phantom{.}\hat{2}}&=&\frac{1}{2}\Big(2h^{\phantom{.}\hat{0}}_2+h^{\phantom{.}\hat{0}}_2h^{\phantom{.}\hat{1}}_1-h^{\phantom{.}\hat{0}}_1h^{\phantom{.}\hat{1}}_2
\Big)\Big(1+\frac{h}{2}\Big)^{-1},
\ee
\be \label{eq:h_s}
h&=&\frac{1}{4}\Big(4h^{\phantom{.}\hat{0}}_{0}-2h^{\phantom{.}\hat{0}}_{1}h^{\phantom{.}\hat{1}}_{0}+4h^{\phantom{.}\hat{1}}_{1}+2h^{\phantom{.}\hat{0}}_{0}h^{\phantom{.}\hat{1}}_{1}-2h^{\phantom{.}\hat{0}}_{2}h^{\phantom{.}\hat{2}}_{0}-h^{\phantom{.}\hat{0}}_{2}h^{\phantom{.}\hat{1}}_{1}h^{\phantom{.}\hat{2}}_{0}+h^{\phantom{.}\hat{0}}_{1}h^{\phantom{.}\hat{1}}_{2}h^{\phantom{.}\hat{2}}_{0}
\nn
&&+h^{\phantom{.}\hat{0}}_{2}h^{\phantom{.}\hat{1}}_{0}h^{\phantom{.}\hat{2}}_{1}-2h^{\phantom{.}\hat{1}}_{2}h^{\phantom{.}\hat{2}}_{1}-h^{\phantom{.}\hat{0}}_{0}h^{\phantom{.}\hat{1}}_{2}h^{\phantom{.}\hat{2}}_{1}+4h^{\phantom{.}\hat{2}}_{2}+2h^{\phantom{.}\hat{0}}_{0}h^{\phantom{.}\hat{2}}_{2}-h^{\phantom{.}\hat{0}}_{1}h^{\phantom{.}\hat{1}}_{0}h^{\phantom{.}\hat{2}}_{2}+2h^{\phantom{.}\hat{1}}_{1}h^{\phantom{.}\hat{2}}_{2}+h^{\phantom{.}\hat{0}}_{0}h^{\phantom{.}\hat{1}}_{1}h^{\phantom{.}\hat{2}}_{2}
\Big).
\ee
Using Eq.~(\ref{eq:h0h2_s}) and Eq.~(\ref{eq:h_s}), these terms and their functional derivatives (under Luttinger's gravitational perturbation) are as follows:
\be
h^{2}_{\phantom{.}\hat{0}}(x^2)\Bigg|_{e^{\phantom{.}\hat{0}}_{\mu}=\delta^{\phantom{.}\hat{0}}_{\mu}(1+\phi_g),
e^{\phantom{.}a}_{\mu}=\delta^{\phantom{.}a}_{\mu}}&=&0,
\nn 
h^{2}_{\phantom{.}\hat{0}}(x^2)\Bigg|_{e^{\phantom{.}\hat{0}}_{\mu}=\delta^{\phantom{.}\hat{0}}_{\mu}(1+\phi_g),
e^{\phantom{.}a}_{\mu}=\delta^{\phantom{.}a}_{\mu}}&=&0,
\nn 
h(x^2)\Bigg|_{e^{\phantom{.}\hat{0}}_{\mu}=\delta^{\phantom{.}\hat{0}}_{\mu}(1+\phi_g),
e^{\phantom{.}a}_{\mu}=\delta^{\phantom{.}a}_{\mu}}&=&2\phi_g(x^2),
\nn
\frac{\delta h^{2}_{\phantom{.}\hat{0}}(x^2)}{\delta h^{\phantom{.}\hat{2}}_{0}(x^2)}\Bigg|_{e^{\phantom{.}\hat{0}}_{\mu}=\delta^{\phantom{.}\hat{0}}_{\mu}(1+\phi_g),
e^{\phantom{.}a}_{\mu}=\delta^{\phantom{.}a}_{\mu}}&=&\frac{1}{1+\phi_g(x^2)},
\nn
\frac{\delta h^{2}_{\phantom{.}\hat{2}}(x^2)}{\delta h^{\phantom{.}\hat{2}}_{0}(x^2)}\Bigg|_{e^{\phantom{.}\hat{0}}_{\mu}=\delta^{\phantom{.}\hat{0}}_{\mu}(1+\phi_g),
e^{\phantom{.}a}_{\mu}=\delta^{\phantom{.}a}_{\mu}}&=&0,
\nn
\frac{\delta h(x^2)}{\delta h^{\phantom{.}\hat{2}}_{0}(x^2)}\Bigg|_{e^{\phantom{.}\hat{0}}_{\mu}=\delta^{\phantom{.}\hat{0}}_{\mu}(1+\phi_g),
e^{\phantom{.}a}_{\mu}=\delta^{\phantom{.}a}_{\mu}}&=&0.
\ee
Combined with Eq.~(\ref{eq:zeta_def_s}), the functional derivative of $\zeta(x^2)$ is given by
\be
\frac{\delta\zeta(x^2)}{\delta h^{\phantom{.}\hat{2}}_{0}(x^2)}&=&\Big(1+\frac{h(x^2)}{2}\Big)\frac{\delta}{\delta h^{\phantom{.}\hat{2}}_{0}(x^2)}\Big(-\frac{1}{2}h^{2}_{\phantom{.}\hat{0}}(x^2)\Big)=-(1+\phi_g(x^2))\frac{1}{2+2\phi_g(x^2)}=-\frac{1}{2}.
\ee
This is exactly the result shown in the main text. Using the same approach, it can be easily demonstrated that
\be
\frac{\delta h^{0}_{\phantom{.}\hat{0}}(x^2)}{\delta h^{\phantom{.}\hat{2}}_{0}(x^2)}\Bigg|_{e^{\phantom{.}\hat{0}}_{\mu}=\delta^{\phantom{.}\hat{0}}_{\mu}(1+\phi_g),
e^{\phantom{.}a}_{\mu}=\delta^{\phantom{.}a}_{\mu}}&=&0,
\nn
\frac{\delta h^{0}_{\phantom{.}\hat{1}}(x^2)}{\delta h^{\phantom{.}\hat{2}}_{0}(x^2)}\Bigg|_{e^{\phantom{.}\hat{0}}_{\mu}=\delta^{\phantom{.}\hat{0}}_{\mu}(1+\phi_g),
e^{\phantom{.}a}_{\mu}=\delta^{\phantom{.}a}_{\mu}}&=&0,
\nn
\frac{\delta h^{0}_{\phantom{.}\hat{2}}(x^2)}{\delta h^{\phantom{.}\hat{2}}_{0}(x^2)}\Bigg|_{e^{\phantom{.}\hat{0}}_{\mu}=\delta^{\phantom{.}\hat{0}}_{\mu}(1+\phi_g),
e^{\phantom{.}a}_{\mu}=\delta^{\phantom{.}a}_{\mu}}&=&0,
\ee
which proves that $\delta e^{0}_{\phantom{.}\alpha}/\delta h_0^{\phantom{.}\hat{2}}=0$, for $\alpha=\hat{0},\hat{1},\hat{2}$.


\begin{thebibliography}{32}%

\twocolumngrid 
\makeatletter
\providecommand \@ifxundefined [1]{%
 \@ifx{#1\undefined}
}%
\providecommand \@ifnum [1]{%
 \ifnum #1\expandafter \@firstoftwo
 \else \expandafter \@secondoftwo
 \fi
}%
\providecommand \@ifx [1]{%
 \ifx #1\expandafter \@firstoftwo
 \else \expandafter \@secondoftwo
 \fi
}%
\providecommand \natexlab [1]{#1}%
\providecommand \enquote  [1]{``#1''}%
\providecommand \bibnamefont  [1]{#1}%
\providecommand \bibfnamefont [1]{#1}%
\providecommand \citenamefont [1]{#1}%
\providecommand \href@noop [0]{\@secondoftwo}%
\providecommand \href [0]{\begingroup \@sanitize@url \@href}%
\providecommand \@href[1]{\@@startlink{#1}\@@href}%
\providecommand \@@href[1]{\endgroup#1\@@endlink}%
\providecommand \@sanitize@url [0]{\catcode `\\12\catcode `\$12\catcode `\&12\catcode `\#12\catcode `\^12\catcode `\_12\catcode `\%12\relax}%
\providecommand \@@startlink[1]{}%
\providecommand \@@endlink[0]{}%
\providecommand \url  [0]{\begingroup\@sanitize@url \@url }%
\providecommand \@url [1]{\endgroup\@href {#1}{\urlprefix }}%
\providecommand \urlprefix  [0]{URL }%
\providecommand \Eprint [0]{\href }%
\providecommand \doibase [0]{https://doi.org/}%
\providecommand \selectlanguage [0]{\@gobble}%
\providecommand \bibinfo  [0]{\@secondoftwo}%
\providecommand \bibfield  [0]{\@secondoftwo}%
\providecommand \translation [1]{[#1]}%
\providecommand \BibitemOpen [0]{}%
\providecommand \bibitemStop [0]{}%
\providecommand \bibitemNoStop [0]{.\EOS\space}%
\providecommand \EOS [0]{\spacefactor3000\relax}%
\providecommand \BibitemShut  [1]{\csname bibitem#1\endcsname}%
\let\auto@bib@innerbib\@empty
\bibitem [{\citenamefont {Ideue}\ et~al.(2017)\citenamefont {Ideue}, \citenamefont {Kurumaji}, \citenamefont {Ishiwata},\ and\ \citenamefont {Tokura}}]{ideue2017giant}%
  \BibitemOpen
  \bibfield  {author} {\bibinfo {author} {\bibfnamefont {T.}~\bibnamefont {Ideue}}, \bibinfo {author} {\bibfnamefont {T.}~\bibnamefont {Kurumaji}}, \bibinfo {author} {\bibfnamefont {S.}~\bibnamefont {Ishiwata}},\ and\ \bibinfo {author} {\bibfnamefont {Y.}~\bibnamefont {Tokura}},\ }\bibfield  {title} {\bibinfo {title} {Giant thermal hall effect in multiferroics},\ }\href@noop {} {\bibfield  {journal} {\bibinfo  {journal} {Nature materials}\ }\textbf {\bibinfo {volume} {16}},\ \bibinfo {pages} {797} (\bibinfo {year} {2017})}\BibitemShut {NoStop}%
\bibitem [{\citenamefont {Shimizu}\ et~al.(2015)\citenamefont {Shimizu}, \citenamefont {Yamakage},\ and\ \citenamefont {Nomura}}]{shimizu2015quantum}%
  \BibitemOpen
  \bibfield  {author} {\bibinfo {author} {\bibfnamefont {Y.}~\bibnamefont {Shimizu}}, \bibinfo {author} {\bibfnamefont {A.}~\bibnamefont {Yamakage}},\ and\ \bibinfo {author} {\bibfnamefont {K.}~\bibnamefont {Nomura}},\ }\bibfield  {title} {\bibinfo {title} {Quantum thermal hall effect of majorana fermions on the surface of superconducting topological insulators},\ }\href@noop {} {\bibfield  {journal} {\bibinfo  {journal} {Physical Review B}\ }\textbf {\bibinfo {volume} {91}},\ \bibinfo {pages} {195139} (\bibinfo {year} {2015})}\BibitemShut {NoStop}%
\bibitem [{\citenamefont {Katsura}\ et~al.(2010)\citenamefont {Katsura}, \citenamefont {Nagaosa},\ and\ \citenamefont {Lee}}]{katsura2010theory}%
  \BibitemOpen
  \bibfield  {author} {\bibinfo {author} {\bibfnamefont {H.}~\bibnamefont {Katsura}}, \bibinfo {author} {\bibfnamefont {N.}~\bibnamefont {Nagaosa}},\ and\ \bibinfo {author} {\bibfnamefont {P.~A.}\ \bibnamefont {Lee}},\ }\bibfield  {title} {\bibinfo {title} {Theory of the thermal hall effect in quantum magnets},\ }\href@noop {} {\bibfield  {journal} {\bibinfo  {journal} {Physical review letters}\ }\textbf {\bibinfo {volume} {104}},\ \bibinfo {pages} {066403} (\bibinfo {year} {2010})}\BibitemShut {NoStop}%
\bibitem [{\citenamefont {Zhuo}\ et~al.(2021)\citenamefont {Zhuo}, \citenamefont {Li}, \citenamefont {Manchon} et~al.}]{zhuo2021topological}%
  \BibitemOpen
  \bibfield  {author} {\bibinfo {author} {\bibfnamefont {F.}~\bibnamefont {Zhuo}}, \bibinfo {author} {\bibfnamefont {H.}~\bibnamefont {Li}}, \bibinfo {author} {\bibfnamefont {A.}~\bibnamefont {Manchon}}, et~al.,\ }\bibfield  {title} {\bibinfo {title} {Topological phase transition and thermal hall effect in kagome ferromagnets},\ }\href@noop {} {\bibfield  {journal} {\bibinfo  {journal} {Physical Review B}\ }\textbf {\bibinfo {volume} {104}},\ \bibinfo {pages} {144422} (\bibinfo {year} {2021})}\BibitemShut {NoStop}%
\bibitem [{\citenamefont {Kawano}\ and\ \citenamefont {Hotta}(2019)}]{kawano2019thermal}%
  \BibitemOpen
  \bibfield  {author} {\bibinfo {author} {\bibfnamefont {M.}~\bibnamefont {Kawano}}\ and\ \bibinfo {author} {\bibfnamefont {C.}~\bibnamefont {Hotta}},\ }\bibfield  {title} {\bibinfo {title} {Thermal hall effect and topological edge states in a square-lattice antiferromagnet},\ }\href@noop {} {\bibfield  {journal} {\bibinfo  {journal} {Physical Review B}\ }\textbf {\bibinfo {volume} {99}},\ \bibinfo {pages} {054422} (\bibinfo {year} {2019})}\BibitemShut {NoStop}%
\bibitem [{\citenamefont {Owerre}(2017)}]{owerre2017topological}%
  \BibitemOpen
  \bibfield  {author} {\bibinfo {author} {\bibfnamefont {S.}~\bibnamefont {Owerre}},\ }\bibfield  {title} {\bibinfo {title} {Topological thermal hall effect in frustrated kagome antiferromagnets},\ }\href@noop {} {\bibfield  {journal} {\bibinfo  {journal} {Physical Review B}\ }\textbf {\bibinfo {volume} {95}},\ \bibinfo {pages} {014422} (\bibinfo {year} {2017})}\BibitemShut {NoStop}%
\bibitem [{\citenamefont {Kasahara}\ et~al.(2018{\natexlab{a}})\citenamefont {Kasahara}, \citenamefont {Sugii}, \citenamefont {Ohnishi}, \citenamefont {Shimozawa}, \citenamefont {Yamashita}, \citenamefont {Kurita}, \citenamefont {Tanaka}, \citenamefont {Nasu}, \citenamefont {Motome}, \citenamefont {Shibauchi} et~al.}]{kasahara2018unusual}%
  \BibitemOpen
  \bibfield  {author} {\bibinfo {author} {\bibfnamefont {Y.}~\bibnamefont {Kasahara}}, \bibinfo {author} {\bibfnamefont {K.}~\bibnamefont {Sugii}}, \bibinfo {author} {\bibfnamefont {T.}~\bibnamefont {Ohnishi}}, \bibinfo {author} {\bibfnamefont {M.}~\bibnamefont {Shimozawa}}, \bibinfo {author} {\bibfnamefont {M.}~\bibnamefont {Yamashita}}, \bibinfo {author} {\bibfnamefont {N.}~\bibnamefont {Kurita}}, \bibinfo {author} {\bibfnamefont {H.}~\bibnamefont {Tanaka}}, \bibinfo {author} {\bibfnamefont {J.}~\bibnamefont {Nasu}}, \bibinfo {author} {\bibfnamefont {Y.}~\bibnamefont {Motome}}, \bibinfo {author} {\bibfnamefont {T.}~\bibnamefont {Shibauchi}}, et~al.,\ }\bibfield  {title} {\bibinfo {title} {Unusual thermal hall effect in a kitaev spin liquid candidate $\alpha$- rucl 3},\ }\href@noop {} {\bibfield  {journal} {\bibinfo  {journal} {Physical review letters}\ }\textbf {\bibinfo {volume} {120}},\ \bibinfo {pages} {217205} (\bibinfo {year} {2018}{\natexlab{a}})}\BibitemShut {NoStop}%
\bibitem [{\citenamefont {Kasahara}\ et~al.(2018{\natexlab{b}})\citenamefont {Kasahara}, \citenamefont {Ohnishi}, \citenamefont {Mizukami}, \citenamefont {Tanaka}, \citenamefont {Ma}, \citenamefont {Sugii}, \citenamefont {Kurita}, \citenamefont {Tanaka}, \citenamefont {Nasu}, \citenamefont {Motome} et~al.}]{kasahara2018majorana}%
  \BibitemOpen
  \bibfield  {author} {\bibinfo {author} {\bibfnamefont {Y.}~\bibnamefont {Kasahara}}, \bibinfo {author} {\bibfnamefont {T.}~\bibnamefont {Ohnishi}}, \bibinfo {author} {\bibfnamefont {Y.}~\bibnamefont {Mizukami}}, \bibinfo {author} {\bibfnamefont {O.}~\bibnamefont {Tanaka}}, \bibinfo {author} {\bibfnamefont {S.}~\bibnamefont {Ma}}, \bibinfo {author} {\bibfnamefont {K.}~\bibnamefont {Sugii}}, \bibinfo {author} {\bibfnamefont {N.}~\bibnamefont {Kurita}}, \bibinfo {author} {\bibfnamefont {H.}~\bibnamefont {Tanaka}}, \bibinfo {author} {\bibfnamefont {J.}~\bibnamefont {Nasu}}, \bibinfo {author} {\bibfnamefont {Y.}~\bibnamefont {Motome}}, et~al.,\ }\bibfield  {title} {\bibinfo {title} {Majorana quantization and half-integer thermal quantum hall effect in a kitaev spin liquid},\ }\href@noop {} {\bibfield  {journal} {\bibinfo  {journal} {Nature}\ }\textbf {\bibinfo {volume} {559}},\ \bibinfo {pages} {227} (\bibinfo {year} {2018}{\natexlab{b}})}\BibitemShut {NoStop}%
\bibitem [{\citenamefont {Teng}\ et~al.(2020)\citenamefont {Teng}, \citenamefont {Zhang}, \citenamefont {Samajdar}, \citenamefont {Scheurer},\ and\ \citenamefont {Sachdev}}]{teng2020unquantized}%
  \BibitemOpen
  \bibfield  {author} {\bibinfo {author} {\bibfnamefont {Y.}~\bibnamefont {Teng}}, \bibinfo {author} {\bibfnamefont {Y.}~\bibnamefont {Zhang}}, \bibinfo {author} {\bibfnamefont {R.}~\bibnamefont {Samajdar}}, \bibinfo {author} {\bibfnamefont {M.~S.}\ \bibnamefont {Scheurer}},\ and\ \bibinfo {author} {\bibfnamefont {S.}~\bibnamefont {Sachdev}},\ }\bibfield  {title} {\bibinfo {title} {Unquantized thermal hall effect in quantum spin liquids with spinon fermi surfaces},\ }\href@noop {} {\bibfield  {journal} {\bibinfo  {journal} {Physical Review Research}\ }\textbf {\bibinfo {volume} {2}},\ \bibinfo {pages} {033283} (\bibinfo {year} {2020})}\BibitemShut {NoStop}%
\bibitem [{\citenamefont {Uehara}\ et~al.(2022)\citenamefont {Uehara}, \citenamefont {Ohtsuki}, \citenamefont {Udagawa}, \citenamefont {Nakatsuji},\ and\ \citenamefont {Machida}}]{uehara2022phonon}%
  \BibitemOpen
  \bibfield  {author} {\bibinfo {author} {\bibfnamefont {T.}~\bibnamefont {Uehara}}, \bibinfo {author} {\bibfnamefont {T.}~\bibnamefont {Ohtsuki}}, \bibinfo {author} {\bibfnamefont {M.}~\bibnamefont {Udagawa}}, \bibinfo {author} {\bibfnamefont {S.}~\bibnamefont {Nakatsuji}},\ and\ \bibinfo {author} {\bibfnamefont {Y.}~\bibnamefont {Machida}},\ }\bibfield  {title} {\bibinfo {title} {Phonon thermal hall effect in a metallic spin ice},\ }\href@noop {} {\bibfield  {journal} {\bibinfo  {journal} {Nature Communications}\ }\textbf {\bibinfo {volume} {13}},\ \bibinfo {pages} {4604} (\bibinfo {year} {2022})}\BibitemShut {NoStop}%
\bibitem [{\citenamefont {Auerbach}(2019)}]{auerbach2019equilibrium}%
  \BibitemOpen
  \bibfield  {author} {\bibinfo {author} {\bibfnamefont {A.}~\bibnamefont {Auerbach}},\ }\bibfield  {title} {\bibinfo {title} {Equilibrium formulae for transverse magnetotransport of strongly correlated metals},\ }\href@noop {} {\bibfield  {journal} {\bibinfo  {journal} {Physical Review B}\ }\textbf {\bibinfo {volume} {99}},\ \bibinfo {pages} {115115} (\bibinfo {year} {2019})}\BibitemShut {NoStop}%
\bibitem [{\citenamefont {Luttinger}(1964)}]{luttinger1964theory}%
  \BibitemOpen
  \bibfield  {author} {\bibinfo {author} {\bibfnamefont {J.}~\bibnamefont {Luttinger}},\ }\bibfield  {title} {\bibinfo {title} {Theory of thermal transport coefficients},\ }\href@noop {} {\bibfield  {journal} {\bibinfo  {journal} {Physical Review}\ }\textbf {\bibinfo {volume} {135}},\ \bibinfo {pages} {A1505} (\bibinfo {year} {1964})}\BibitemShut {NoStop}%
\bibitem [{\citenamefont {Cooper}\ et~al.(1997)\citenamefont {Cooper}, \citenamefont {Halperin},\ and\ \citenamefont {Ruzin}}]{cooper1997thermoelectric}%
  \BibitemOpen
  \bibfield  {author} {\bibinfo {author} {\bibfnamefont {N.}~\bibnamefont {Cooper}}, \bibinfo {author} {\bibfnamefont {B.}~\bibnamefont {Halperin}},\ and\ \bibinfo {author} {\bibfnamefont {I.}~\bibnamefont {Ruzin}},\ }\bibfield  {title} {\bibinfo {title} {Thermoelectric response of an interacting two-dimensional electron gas in a quantizing magnetic field},\ }\href@noop {} {\bibfield  {journal} {\bibinfo  {journal} {Physical Review B}\ }\textbf {\bibinfo {volume} {55}},\ \bibinfo {pages} {2344} (\bibinfo {year} {1997})}\BibitemShut {NoStop}%
\bibitem [{\citenamefont {Qin}\ et~al.(2011)\citenamefont {Qin}, \citenamefont {Niu},\ and\ \citenamefont {Shi}}]{qin2011energy}%
  \BibitemOpen
  \bibfield  {author} {\bibinfo {author} {\bibfnamefont {T.}~\bibnamefont {Qin}}, \bibinfo {author} {\bibfnamefont {Q.}~\bibnamefont {Niu}},\ and\ \bibinfo {author} {\bibfnamefont {J.}~\bibnamefont {Shi}},\ }\bibfield  {title} {\bibinfo {title} {Energy magnetization and the thermal hall effect},\ }\href@noop {} {\bibfield  {journal} {\bibinfo  {journal} {Physical review letters}\ }\textbf {\bibinfo {volume} {107}},\ \bibinfo {pages} {236601} (\bibinfo {year} {2011})}\BibitemShut {NoStop}%
\bibitem [{\citenamefont {Shitade}(2014)}]{shitade2014heat}%
  \BibitemOpen
  \bibfield  {author} {\bibinfo {author} {\bibfnamefont {A.}~\bibnamefont {Shitade}},\ }\bibfield  {title} {\bibinfo {title} {Heat transport as torsional responses and keldysh formalism in a curved spacetime},\ }\href@noop {} {\bibfield  {journal} {\bibinfo  {journal} {Progress of Theoretical and Experimental Physics}\ }\textbf {\bibinfo {volume} {2014}},\ \bibinfo {pages} {123I01} (\bibinfo {year} {2014})}\BibitemShut {NoStop}%
\bibitem [{\citenamefont {Nakai}\ et~al.(2016)\citenamefont {Nakai}, \citenamefont {Ryu},\ and\ \citenamefont {Nomura}}]{nakai2016finite}%
  \BibitemOpen
  \bibfield  {author} {\bibinfo {author} {\bibfnamefont {R.}~\bibnamefont {Nakai}}, \bibinfo {author} {\bibfnamefont {S.}~\bibnamefont {Ryu}},\ and\ \bibinfo {author} {\bibfnamefont {K.}~\bibnamefont {Nomura}},\ }\bibfield  {title} {\bibinfo {title} {Finite-temperature effective boundary theory of the quantized thermal hall effect},\ }\href@noop {} {\bibfield  {journal} {\bibinfo  {journal} {New Journal of Physics}\ }\textbf {\bibinfo {volume} {18}},\ \bibinfo {pages} {023038} (\bibinfo {year} {2016})}\BibitemShut {NoStop}%
\bibitem [{\citenamefont {Nakai}\ et~al.(2017)\citenamefont {Nakai}, \citenamefont {Ryu},\ and\ \citenamefont {Nomura}}]{nakai2017laughlin}%
  \BibitemOpen
  \bibfield  {author} {\bibinfo {author} {\bibfnamefont {R.}~\bibnamefont {Nakai}}, \bibinfo {author} {\bibfnamefont {S.}~\bibnamefont {Ryu}},\ and\ \bibinfo {author} {\bibfnamefont {K.}~\bibnamefont {Nomura}},\ }\bibfield  {title} {\bibinfo {title} {Laughlin's argument for the quantized thermal hall effect},\ }\href@noop {} {\bibfield  {journal} {\bibinfo  {journal} {Physical Review B}\ }\textbf {\bibinfo {volume} {95}},\ \bibinfo {pages} {165405} (\bibinfo {year} {2017})}\BibitemShut {NoStop}%
\bibitem [{\citenamefont {Stone}(2012)}]{stone2012gravitational}%
  \BibitemOpen
  \bibfield  {author} {\bibinfo {author} {\bibfnamefont {M.}~\bibnamefont {Stone}},\ }\bibfield  {title} {\bibinfo {title} {Gravitational anomalies and thermal hall effect in topological insulators},\ }\href@noop {} {\bibfield  {journal} {\bibinfo  {journal} {Physical Review B}\ }\textbf {\bibinfo {volume} {85}},\ \bibinfo {pages} {184503} (\bibinfo {year} {2012})}\BibitemShut {NoStop}%
\bibitem [{\citenamefont {Bradlyn}\ and\ \citenamefont {Read}(2015)}]{bradlyn2015low}%
  \BibitemOpen
  \bibfield  {author} {\bibinfo {author} {\bibfnamefont {B.}~\bibnamefont {Bradlyn}}\ and\ \bibinfo {author} {\bibfnamefont {N.}~\bibnamefont {Read}},\ }\bibfield  {title} {\bibinfo {title} {Low-energy effective theory in the bulk for transport in a topological phase},\ }\href@noop {} {\bibfield  {journal} {\bibinfo  {journal} {Physical Review B}\ }\textbf {\bibinfo {volume} {91}},\ \bibinfo {pages} {125303} (\bibinfo {year} {2015})}\BibitemShut {NoStop}%
\bibitem [{\citenamefont {Gromov}\ and\ \citenamefont {Abanov}(2015)}]{gromov2015thermal}%
  \BibitemOpen
  \bibfield  {author} {\bibinfo {author} {\bibfnamefont {A.}~\bibnamefont {Gromov}}\ and\ \bibinfo {author} {\bibfnamefont {A.~G.}\ \bibnamefont {Abanov}},\ }\bibfield  {title} {\bibinfo {title} {Thermal hall effect and geometry with torsion},\ }\href@noop {} {\bibfield  {journal} {\bibinfo  {journal} {Physical review letters}\ }\textbf {\bibinfo {volume} {114}},\ \bibinfo {pages} {016802} (\bibinfo {year} {2015})}\BibitemShut {NoStop}%
\bibitem [{\citenamefont {Vinkler-Aviv}(2019)}]{vinkler2019bulk}%
  \BibitemOpen
  \bibfield  {author} {\bibinfo {author} {\bibfnamefont {Y.}~\bibnamefont {Vinkler-Aviv}},\ }\bibfield  {title} {\bibinfo {title} {Bulk thermal transport coefficients in a quantum hall system and the fundamental difference between thermal and charge response},\ }\href@noop {} {\bibfield  {journal} {\bibinfo  {journal} {Physical Review B}\ }\textbf {\bibinfo {volume} {100}},\ \bibinfo {pages} {041106} (\bibinfo {year} {2019})}\BibitemShut {NoStop}%
\bibitem [{\citenamefont {Huang}\ et~al.(2022)\citenamefont {Huang}, \citenamefont {Han},\ and\ \citenamefont {Sun}}]{huang2022torsion}%
  \BibitemOpen
  \bibfield  {author} {\bibinfo {author} {\bibfnamefont {Z.-M.}\ \bibnamefont {Huang}}, \bibinfo {author} {\bibfnamefont {B.}~\bibnamefont {Han}},\ and\ \bibinfo {author} {\bibfnamefont {X.-Q.}\ \bibnamefont {Sun}},\ }\bibfield  {title} {\bibinfo {title} {Torsion, energy magnetization, and thermal hall effect},\ }\href@noop {} {\bibfield  {journal} {\bibinfo  {journal} {Physical Review B}\ }\textbf {\bibinfo {volume} {105}},\ \bibinfo {pages} {085104} (\bibinfo {year} {2022})}\BibitemShut {NoStop}%
\bibitem [{Note1()}]{Note1}%
  \BibitemOpen
  \bibinfo {note} {Note that the boundary energy current is not dependent on time, since the boundary free energy~(\ref {eq:free_energy_result}) and responses derived from it are static. Eq.~(\ref {eq:bulk_current}) is obtained by assuming that a steady state is achieved. Therefore, the divergence of the energy current must be equal to zero. At the boundary, the current has two contributions: the bulk one $j^{1}_E(x^1,x^2)$ flowing in the $x^1$ direction, and the boundary one $j^{2}_E(x^1,x^2) = j^{\protect \text {bdry}}_E(x^2) \delta (x^1)$ flowing along the boundary in the direction $x^2$ and localized at $x^1=0$. Integrating the equation $\partial _\mu j^{\mu }_E = 0$ over $x^1$ across the boundary and using that, by symmetry, $j^{1}_E(x^1=-0)=-j^1_E(x^1=+0)$, one readily obtains Eq.~(\ref {eq:bulk_current}).}\BibitemShut {Stop}%
\bibitem [{\citenamefont {Nomura}\ et~al.(2012)\citenamefont {Nomura}, \citenamefont {Ryu}, \citenamefont {Furusaki},\ and\ \citenamefont {Nagaosa}}]{nomura2012cross}%
  \BibitemOpen
  \bibfield  {author} {\bibinfo {author} {\bibfnamefont {K.}~\bibnamefont {Nomura}}, \bibinfo {author} {\bibfnamefont {S.}~\bibnamefont {Ryu}}, \bibinfo {author} {\bibfnamefont {A.}~\bibnamefont {Furusaki}},\ and\ \bibinfo {author} {\bibfnamefont {N.}~\bibnamefont {Nagaosa}},\ }\bibfield  {title} {\bibinfo {title} {Cross-correlated responses of topological superconductors and superfluids},\ }\href@noop {} {\bibfield  {journal} {\bibinfo  {journal} {Physical review letters}\ }\textbf {\bibinfo {volume} {108}},\ \bibinfo {pages} {026802} (\bibinfo {year} {2012})}\BibitemShut {NoStop}%
\bibitem [{Note2()}]{Note2}%
  \BibitemOpen
  \bibinfo {note} {In this paper, we have adopted a form of the action that differs from the one in Ref.~\cite {nakai2016finite}. Specifically, we have added total derivatives to the Lagrangian to make it Hermitian. However, this difference does not alter our final conclusions. Based on the action from Refs.~\cite {nakai2016finite}, the boundary free energy obtained is the same as the one presented in Eq.~(\ref {eq:free_energy_result}) of the main text, up to a constant that is independent of temperature.}\BibitemShut {Stop}%
\bibitem [{\citenamefont {Carroll}(2019)}]{carroll2019spacetime}%
  \BibitemOpen
  \bibfield  {author} {\bibinfo {author} {\bibfnamefont {S.~M.}\ \bibnamefont {Carroll}},\ }\href@noop {} {\bibinfo {title} {Spacetime and geometry}}\ (\bibinfo  {publisher} {Cambridge University Press},\ \bibinfo {year} {2019})\BibitemShut {NoStop}%
\bibitem [{sup()}]{supplemental_material}%
  \BibitemOpen
  \href@noop {} {\bibinfo {title} {See supplemental material for details of the derivations.}}\BibitemShut {Stop}%
\bibitem [{\citenamefont {Stoof}\ et~al.(2009)\citenamefont {Stoof}, \citenamefont {Gubbels},\ and\ \citenamefont {Dickerscheid}}]{stoof2009ultracold}%
  \BibitemOpen
  \bibfield  {author} {\bibinfo {author} {\bibfnamefont {H.~T.}\ \bibnamefont {Stoof}}, \bibinfo {author} {\bibfnamefont {K.~B.}\ \bibnamefont {Gubbels}},\ and\ \bibinfo {author} {\bibfnamefont {D.}~\bibnamefont {Dickerscheid}},\ }\href@noop {} {\bibinfo {title} {Ultracold quantum fields}}\ (\bibinfo  {publisher} {Springer},\ \bibinfo {year} {2009})\BibitemShut {NoStop}%
\bibitem [{Note3()}]{Note3}%
  \BibitemOpen
  \bibinfo {note} {Since we neglected temperature-independent terms in our derivation, this form of the boundary free energy is only valid up to a constant which is independent of temperature.}\BibitemShut {Stop}%
\bibitem [{\citenamefont {Zhang}\ et~al.(2020)\citenamefont {Zhang}, \citenamefont {Gao},\ and\ \citenamefont {Xiao}}]{zhang2020thermodynamics}%
  \BibitemOpen
  \bibfield  {author} {\bibinfo {author} {\bibfnamefont {Y.}~\bibnamefont {Zhang}}, \bibinfo {author} {\bibfnamefont {Y.}~\bibnamefont {Gao}},\ and\ \bibinfo {author} {\bibfnamefont {D.}~\bibnamefont {Xiao}},\ }\bibfield  {title} {\bibinfo {title} {Thermodynamics of energy magnetization},\ }\href@noop {} {\bibfield  {journal} {\bibinfo  {journal} {Physical Review B}\ }\textbf {\bibinfo {volume} {102}},\ \bibinfo {pages} {235161} (\bibinfo {year} {2020})}\BibitemShut {NoStop}%
\bibitem [{Note4()}]{Note4}%
  \BibitemOpen
  \bibinfo {note} {Note indeed that, according to Eq.~(\ref {eq:free_energy_s}), to obtain $F$ we have to multiply by $\beta ^{-1}$}\BibitemShut {NoStop}%
\bibitem [{\citenamefont {Hertz}\ and\ \citenamefont {Klenin}(1974)}]{hertz1974fluctuations}%
  \BibitemOpen
  \bibfield  {author} {\bibinfo {author} {\bibfnamefont {J.}~\bibnamefont {Hertz}}\ and\ \bibinfo {author} {\bibfnamefont {M.}~\bibnamefont {Klenin}},\ }\bibfield  {title} {\bibinfo {title} {Fluctuations in itinerant-electron paramagnets},\ }\href@noop {} {\bibfield  {journal} {\bibinfo  {journal} {Physical Review B}\ }\textbf {\bibinfo {volume} {10}},\ \bibinfo {pages} {1084} (\bibinfo {year} {1974})}\BibitemShut {NoStop}%
\end{thebibliography}
\end{document}